\documentclass[aps,pra,showpacs,twoside,twocolumn,10pt,nofootinbib]{revtex4-2}
\usepackage[utf8]{inputenc}
\usepackage[sc,osf]{mathpazo}
\usepackage{amsmath}
\usepackage[T1]{fontenc}
\usepackage{latexsym}
\usepackage{amssymb}
\usepackage[colorlinks=true,citecolor=blue,urlcolor=blue]{hyperref}
\usepackage{color}
\usepackage{graphics,epstopdf}
\usepackage{soul}
\usepackage[demo]{graphicx}
\usepackage{capt-of}
\usepackage{lipsum}
\usepackage{adjustbox}
\usepackage[normalem]{ulem}
\usepackage[table,xcdraw]{xcolor}
\usepackage[dvipsnames]{xcolor}
\usepackage{braket}
\usepackage{physics}
\usepackage{ragged2e}
\usepackage{mathtools}

\usepackage{comment}
\usepackage{orcidlink}
\usepackage[font=small,labelfont=bf,justification=justified,singlelinecheck=false]{caption}
\usepackage{subfigure}  
\usepackage{bm}
\usepackage{bbm}
\usepackage{braket}
\usepackage{natbib}
\usepackage[table,xcdraw]{xcolor}
\usepackage[dvipsnames]{xcolor}
\usepackage{cancel}
\usepackage{textcomp}

\sethlcolor{yellow}

\newcommand{\Rivu}[1]{{\textcolor{teal} {\it{[Note (Rivu): #1]}}}} 

\setcitestyle{maxbibnames=5,sort&compress}

\begin{document}

\title{Surpassing Gaussian optimality in multiparameter estimation with indefinite causal order}

\author{Sudipta Das $^{1}$\,\orcidlink{0000-0002-9589-2042}} 
\author{Rivu Gupta $^{2}$\,\orcidlink{0000-0002-0254-2028}}
\author{Aditi Sen(De) $^{3, 4}$\,\orcidlink{0000-0003-1693-0440}}
\author{Himadri Shekhar Dhar $^{1, 5}$\,\orcidlink{0000-0002-5877-3415}}
\affiliation{$^{1}$ Department of Physics, Indian Institute of Technology Bombay, Powai, Mumbai, Maharashtra 400076, India}
\affiliation{$^{2}$ Dipartimento di Fisica ``Aldo Pontremoli'', Universit{\'a} degli Studi di Milano, I-20133 Milano, Italy}
\affiliation{$^{3}$ Harish-Chandra Research Institute, Chhatnag Road, Jhunsi, Prayagraj - $211019$, India}
\affiliation{$^{4}$ Homi Bhabha National Institute, Training School Complex, Anushakti Nagar, Mumbai $400 094$, India}
\affiliation{$^{5}$ Centre of Excellence in Quantum Information, Computation, Science and Technology,
Indian Institute of Technology Bombay, Powai, Mumbai, Maharashtra 400076, India}



\begin{abstract}



We identify single-mode Gaussian probes, generated by displacement and squeezing operations on the vacuum state, which are optimal for the simultaneous estimation of displacement and squeezing operations in continuous-variable quantum systems. Importantly, our results reveal that the best precision at a fixed energy is achieved not by an experimentally costly squeezing resource, but rather by redirecting some of the energy towards displacement, thus allowing for more resource-effective operations.
Furthermore, introducing indefinite causal order (ICO) in either the probe preparation or parameter encoding step can surpass the Gaussian precision bound, even though the optimal Gaussian probe state is agnostic to the ordering of the operations. Specifically, we observe that odd-parity superpositions of the two definite orders can enhance precision over optimal Gaussian probes in specific parameter regimes. Further, the observed advantage cannot be attributed solely to non-Gaussianity, as quantified by the relative entropy of non-Gaussianity, highlighting ICO as an independent resource for enhancing multiparameter estimation.


\end{abstract}

\maketitle

\section{Introduction}
\label{sec:intro}

Quantum metrology~\cite{Paris_IJQI_2009_metrology-overview}, a major pillar of quantum technologies, addresses the problem of inferring the value of a physical quantity that cannot be accessed directly~\cite{Ghosh_arXiv_2026_metrology_review}. At its core, it comprises the design of optimal estimation techniques~\cite{Liu_AQT_2022_optimal_metrology, Liu_PRL_2023_optimal_metrology_strategy, Liu_PRA_2025_optimal_local_measurement} to ensure the highest possible precision
allowed by quantum mechanics while remaining experimentally feasible~\cite{Toth_JPA_2014_metrology_information}.
In the single-parameter setting, one seeks to estimate an individual unknown parameter, such as a frequency~\cite{Bollinger_PRA_1996_frequency_estimation, Huelga_PRL_1997_frequency_estimation}, or a quantum phase~\cite{Monras_PRA_2006_phase_estimation, Oh_NPJ_2019, Rodriguez_Quantum_2024, Meher_PRA_2024, Zhao_Optica_2024, Qin_PRA_2025, Park_arXiv_2025_phase-estimation_non-gaussian} while multiparameter estimation simultaneously infers several unknown parameters, with the goal of minimizing the overall estimation error~\cite{Albarelli_PLA_2020_multiparameter_perspective, Szczykulska_AP_2016_multiparameter_metrology, Liu_JPA_2020_QFIM_multiparameter}. Paradigmatic instances include joint estimation of unitary and loss parameters~\cite{Genoni_PRA_2013_joint_parameter, Bradshaw_PLA_2017_joint_parameter, Bradshaw_PRA_2018_joint_parameter}, multiple phases~\cite{Humphreys_PRL_2013_multiple_phases, Gagatsos_PRA_2016_multiple_phases}, 
and magnetic fields~\cite{Zhang_SR_2014_multiple_magmetic-fields}. 
In both cases, the optimal estimation technique
may be global~\cite{VanTrees_2004_Global_estimation}, which is independent of the value of the parameter to be estimated, or local~\cite{Helstrom_JSP_1969_attain-SLD}, where the estimator is optimized around a particular parameter value by minimizing its variance~\cite{VanTrees_2004_Global_estimation, Helstrom_JSP_1969_attain-SLD, Holevo_2003_statistical_structure, Helstrom_IEEE_1974_non-commuting_detection, Braunstein_AP_1996_generalized_uncertainty}.

Given the importance and broader applications~\cite{Abbott_PRL_2016_gravitational_waves, Contreras_JMP_2016_Fisher_quantum_geometry, Fiderer_PRXQ_2021_superresolution} of quantum metrology, it is essential to identify scalable and experimentally accessible platforms for its implementation. In this respect, continuous-variable (CV) systems~\cite{Braunstein_RMP_2005_QI_With_CV, Serafini_2017_CV-book, Weedbrook_RMP_2012_Gaussian_QI, Adesso_OSID_2014_CV-review, Walschaers_PRX_2021_non-Gaussian}, which admit a continuous spectrum in an infinite-dimensional Hilbert space, offer distinct advantages in practical quantum metrology~\cite{Pirandola_NP_2018_advances_photonic_sensing, Fadel_RoPP_2025_CV_sensing}. Importantly, quantum information processing tasks using CV systems~\cite{Andersen_Laser_Photonics_Reviews_2010_CV_QI_Processing} have already been experimentally demonstrated in a variety of platforms, including optical and integrated photonic systems~\cite{Clark_Nature_Photonics_2026, Lenzini_Science_Advns_2018}, superconducting 
circuits~\cite{gu2017microwave, hillmann2020universal, eriksson2024universal}, and optomechanical systems~\cite{Stannigel_PRL_2012_optomechanical_QI, Zhou_EPJD_2025_optomechanical_sensing, Huang_FR_2026_optomechanical_entanglement}. From a metrological perspective, single-mode Gaussian states, especially squeezed states~\cite{Gessner_NC_2020_multiparameter_squeezing, Patra_arXiv_2026_beating_noise_squeezing, Manju_arxIV_2026_squeezing_catalyst}, have been shown to attain Heisenberg scaling~\cite{Grochowski_PRL_2025_phase-insensitive_non-Gaussian_metrology, Fadel_RoPP_2025_CV_sensing, Gordillo-Hachuel_arXiv_2026_metrology_higher-order_squeezing} with local measurements~\cite{Wang_PRA_2020_cv_graph-state_homodyne}, thereby surpassing the standard quantum limit. Moreover, the scalability~\cite{van-Loock_PRA_2007_Gaussian-cluster, Menicucci_PRA_2007_ultracompact-CV-cluster, Menicucci_PRA_2011_temporal-mode-CV-cluster, Yokoyama_NP_2013_large-scale-CV-cluster} and controllability~\cite{PhysRevLett.82.1784, krastanov2015universal} of CV systems enable the preparation of large-scale multimode states, making them useful for entanglement-assisted metrology~\cite{Demlowicz_prl_2014_entanglement_noise_metrology, Huang_PRA_2020_auxiliary_metrology}, and  interferometry~\cite{Matsubara_NJP_2019_multimode_interferometry}. 
CV systems also provide a natural setting for multiparameter estimation, with applications ranging from thermometry~\cite{Chattopadhyay_PRA_2025_thermometry}, and Gaussian unitary processes~\cite{Oh_NJP_2020_Gaussian_unitary}, to mitigating the sloppiness in joint estimation~\cite{Sharma_arXiv_2025_mitigating_sloppiness, Manju_arXiv_2025_scrambling_sloppiness}.

In this work, we 
analyze the multiparameter estimation of two exemplary Gaussian unitary processes, \textit{displacement} and \textit{squeezing}~\cite{Ferraro_Bib_2005_Gaussian-CV-review, Adesso_OSID_2014_CV-review}. Despite their ease of implementation in contemporary optics~\cite{Keller_OE_2008_easy_squeezing}, accurate estimation of these quantities is imperative for practical realization of several quantum information protocols~\cite{Braunstein_PRL_1998_CV_teleportation, Braunstein_PRA_2000_CV_dense-coding}, 
especially in the presence of experimental
imperfections such as noise in lasers and parametric amplifiers.
%
While the simultaneous estimation of the displacement and squeezing parameters has been recently addressed using single-mode Gaussian states~\cite{Bressanini_JPA_2024_Gaussian_multiparameter}, the optimal probe state remains unidentified. 
We address this gap by considering a generic pure single-mode Gaussian state at a fixed energy and deriving the optimality condition such that it can furnish the lowest error bound. This leads to two important observations -- firstly, we find that the optimality condition depends on the energy of the probe state, and displaced-squeezed or squeezed-coherent states can both achieve the minimum error bound when this condition is satisfied, independent of the order of the probe-preparation operations;
secondly, since the optimal probe state requires both squeezing and displacement, our results demonstrate that 
instead of expending the entirety of the available energy on squeezing, which has been established to be vital for Heisenberg-limited optical metrology~\cite{Pinel_PRA_2013_single-mode_metrology, Matsubara_NJP_2019_multimode_interferometry}, a part of the energy of the optimal probe is redirected towards displacement, thereby improving precision while reducing the experimentally costly squeezing resource~\cite{Pinel_PRA_2013_single-mode_metrology, Matsubara_NJP_2019_multimode_interferometry}.
We also show that energy acts as a resource, since a more energetic probe state allows for the sensing of greater parameter ranges. 

However, if the optimality condition is not satisfied, the estimation precision at a fixed energy depends on which Gaussian operation during probe preparation is performed first.
This prompts us to ask whether an indefinite order of these operations can enhance metrological performance. In CV systems, such indefinite causal order (ICO) has been shown to aid in phase-space displacement sensing, allowing one to surpass the Heisenberg limit~\cite{Zhao_PRL_2020_ICO_metrology}, with subsequent experimental demonstration~\cite{Yin_2023}.
Using a quantum switch~\cite{Chiribella_PRA_2013_computation_ICO, Procopio_NC_2015_experimental_switch, Wei_PRL_2019_experiment_switch_communication} setup to incorporate ICO among the unitary operations, in both probe preparation and encoding, we demonstrate that superpositions of pure Gaussian states can enhance metrological sensitivity.
In particular, for ICO in probe preparation, we show that the odd-parity state, formed by an equal superposition of the two definite orders, can achieve lower estimation errors than the optimal Gaussian probe over specific parameter regimes. Moreover, ICO in the encoding stage yields a similar parity-dependent behavior, with the odd-parity state again providing the best performance, although the resulting advantage is weaker than that obtained during probe preparation. Furthermore, the behavior of the relative entropy-based non-Gaussianity measure of the ICO probes and encoded states indicates that while non-Gaussianity aids in increasing the precision, it is not the only physical property responsible for the observed metrological benefit.

The paper is structured as follows. In Sec.~\ref{sec:disp_sq_Gaussian}, we derive the optimality condition for Gaussian probes at a fixed energy, where we provide explicit expressions of the error bounds. Having identified the optimal probe state, we analyze the benefits of using indefinite causal order in the probe preparation step in Sec.~\ref{sec:ICO}, while Sec.~\ref{sec:ICO_encoding} comprises the study of ICO in the encoding process. Sec.~\ref{sec:conclu}  includes a summary of our results and possible open directions.

\section{Estimating displacement and squeezing: optimality of pure Gaussian probes}
\label{sec:disp_sq_Gaussian}

In this section, we study how generic pure single-mode Gaussian probes can be employed to estimate displacement and squeezing parameters. Before proceeding to the analysis, let us briefly recall the theory of multiparameter estimation and the figures of merit used to assess the efficiency of the protocol under study. A detailed discussion is provided in Appendix~\ref{app:pre_multiparameter}.

\subsection{A brief recap of multiparameter estimation theory}

In multiparameter quantum estimation, the aim is to determine a set of parameters, $\vec{\theta} = \{\theta_1, \theta_2, \ldots, \theta_m\}$, as accurately as possible. If the parameters correspond to non-commuting generators, it has been established that simultaneous estimation can outperform the strategy of estimating each parameter individually~\cite{Humphreys_PRL_2013_multiple-phase_estimation, Baumgratz_PRL_2016_multidimensional-field_estimation}. We begin with a probe state, $\rho(\vec{\eta})$, characterized by a set of parameters, $\vec{\eta}$, which is independent of the parameters to be estimated. The probe state undergoes evolution through a quantum channel, $\Lambda(\vec{\theta})$, which encodes the information about the parameters into the probe state to yield $\rho(\vec{\theta}; \vec{\eta}) = \Lambda(\vec{\theta})[\rho(\vec{\eta})]$. The encoded state is measured using a POVM, $\hat{\Pi} = \{\hat{\Pi}(x)| \hat{\Pi}(x) \geq 0 ~\forall~ x, \sum_x \hat{\Pi}(x) = \mathbb{I}\}$, whose output statistics are used to estimate the encoded parameters. The mean-square error of the estimated parameters is quantified by the covariance matrix, whose diagonal elements quantify the statistical error for each parameter, while the off-diagonal elements characterize correlations between estimation errors of different parameters. The total error incurred during the process is bounded by the quantum Cramer-Rao bound (CRB)~\cite{Helstrom_PLA_1967_SLD-CRB, Helstrom_IEEE_1968_SLD-CRB}, $C(\vec{\theta};\vec{\eta})$, which can be computed using the quantum Fisher information matrix (QFIM), $\mathcal{Q}(\vec{\theta};\vec{\eta})$~\cite{Liu_JPA_2020_QFIM_multiparameter}, as
\begin{eqnarray}
   &&  C(\vec{\theta}; \vec{\eta}) = \mathrm{Tr}[\mathcal{Q}(\vec{\theta}; \vec{\eta})^{-1}], \\
   && \text{with}~ \mathcal{Q}_{ij}(\vec{\theta}; \vec{\eta}) = \frac{1}{2}\,\mathrm{Tr}\!\left[\rho(\vec{\theta}; \vec{\eta}) \left\{ \hat{\mathcal{L}}_i, \hat{\mathcal{L}}_j \right\}_+ \right].
\label{eq_qfim_element}
\end{eqnarray}
Here, $\{A, B\}_+ = AB + BA$ is the anticommutator, and the operators $\hat{\mathcal{L}}_i$, corresponding to the parameters $\theta_i$, are known as the symmetric logarithmic derivatives (SLD). They are implicitly defined as the solution to~\cite{Helstrom_PLA_1967_SLD-CRB, Braunstein_PRL_1994_attain-CRB, Paris_IJQI_2009_metrology-overview}
\begin{eqnarray}
     \frac{d\rho({\vec{\theta}; \vec{\eta}})}{d\theta_i}=\frac{1}{2}\Big( \hat{\mathcal{L}}_i \rho(\vec{\theta}; \vec{\eta}) + \rho(\vec{\theta}; \vec{\eta})\hat{\mathcal{L}}_i \Big).
    \label{eq_rho_sld}
\end{eqnarray}
For a pure encoded state, the SLD operator can be evaluated directly as
\begin{eqnarray}
    \hat{\mathcal{L}}_i=2 \Big( |\partial_i\psi(\vec{\theta}; \vec{\eta})\rangle \langle \psi(\vec{\theta}; \vec{\eta})|+ |\psi(\vec{\theta}; \vec{\eta})\rangle \langle\partial_i\psi(\vec{\theta}; \vec{\eta})| \Big).
    \label{eq:sld_op}
\end{eqnarray}
While the SLD operators are Hermitian and traceless, the QFIM measured using this operator is real-valued, symmetric, and positive semidefinite.
In multiparameter estimation, even though the SLD-based measure of Cramer-Rao bound (SLD-CRB) does not depend on the choice of the POVM, it cannot be saturated unless the SLDs commute with each other on average, 
in which case one can construct a POVM in their common eigenbasis~\cite{Chang_CP_2026_multiparameter_Gaussian}. However, this is not generally the case in realistic physical setups, 
and from a measurement perspective, the SLD-CRB is not a tight bound, since it cannot
be saturated even in the asymptotic limit. The most fundamental scalar bound for multiparameter estimation is the Holevo-Cramer-Rao bound (HCRB)~\cite{Holevo_JMA_1973_HCRB}, which is tighter than the SLD-CRB by at most a factor of two~\cite{Holevo_2011, Albarelli_PRL_2019, Demkowicz-Dobrzaski_JPA_2020_multiparameter_beyond-Fisher, Zhou_2025}. The HCRB requires asymptotic collective measurements to achieve~\cite{Kahn_CMP_2009_HCRB_asymptotic, Yamagata_AS_2013_HRCB_asymptotic, Yang_CMP_2019_HCRB_asymptotic}, while calculating it requires complicated optimization techniques, and is generally formulated as a semidefinite programming problem for finite~\cite{Albarelli_PRL_2019, Conlon_NPJQI_2021_NCRB_semidefinite}, as well as infinite-dimensional systems~\cite{Bressanini_JPA_2024_Gaussian_multiparameter, Chang_CP_2026_multiparameter_Gaussian}.

\subsection{SLD-CRB for sensing displacement and squeezing using a generic Gaussian state}
\label{subsec:sld-crb_generic}

The multiparameter estimation protocol considered in this work involves determining a complex displacement parameter, $\alpha = \alpha_x + \iota \alpha_p$ with $\alpha_{x(p)} \in \mathbb{R}$, of the displacement operator, $\hat{D}(\alpha) = \exp (\alpha \hat{a}^\dagger - \alpha^* \hat{a})$, and the real part, $r$, of the squeezing parameter, $z = r e^{\iota \varphi} ~\text{with}~ r \in \mathbb{R}^+$, of the squeezing operator, $\hat{S}(z) \equiv \hat{S}(r, \varphi) = \exp \Big( \frac{1}{2}(z \hat{a}^{\dagger 2} - z^* \hat{a}^2) \Big)$. 
Here, $\hat{a}^\dagger$ and $\hat{a}$ denote the creation and annihilation operators respectively, which satisfy $[\hat{a}_j, \hat{a}_k^\dagger] = \delta_{jk}$~\cite{Barnett_2002_Optics-book}, where $j$ and $k$ represent independent modes, $\delta_{jk}$ is the Kronecker delta ($\delta_{jk} = 1 ~\text{if}~ j = k ~\text{and}~0 ~\text{otherwise}$), and $\iota = \sqrt{-1}$ (see Appendix~\ref{app:Gaussian_CV} for a primer on Gaussian CV systems). 

The set of unknown parameters is given by $\vec{\theta} = \{ \alpha_x, \alpha_p, r \}$ and is typically beyond the control of the sensing protocol or experiment. These parameters are imprinted on a probe system through a Gaussian unitary channel, $\Lambda(\vec{\theta})$, consisting of the sequential application of the displacement and the squeezing operators, i.e., $\Lambda(\vec{\theta})=\mathcal{D}(\alpha)\circ\mathcal{S}(z)$, where $\mathcal{D}(\alpha)$ denotes the displacement channel, whose action on a state, $\rho$, is defined as $\mathcal{D}[\rho] = \hat{D} \rho \hat{D}^\dagger$, and similarly for the squeezing channel, $\mathcal{S}(z)$. 
The probe state is taken to be a pure single-mode Gaussian state, obtained by applying the displacement and squeezing operations on vacuum.
Note that the displacement and squeezing unitaries do not commute; hence, the order of their application during probe preparation and encoding also determines the protocol's performance, as quantified by the SLD-CRB. For our purpose, we consider two different configurations of the probe state as $\hat{D}(\beta) \hat{S}(\xi) \ket{0}$ (displaced-squeezed state), and $\hat{S}(\xi) \hat{D}(\beta) \ket{0}$ (squeezed-coherent state), where $\beta = \beta_x + \iota \beta_p$ is
the probe displacement with $\beta_{x(p)} \in \mathbb{R}$, and $\xi = \zeta e^{\iota \kappa}$ is the probe squeezing with $\kappa \in (0, \pi),~\text{being the squeezing angle and}~ \zeta \in \mathbb{R}^+ ~\text{represents the squeezing strength}$.
The states can be written in a compact form as $\ket{\phi(\vec{\eta})}_{DS}$ and $\ket{\phi(\vec{\eta})}_{SD}$, respectively, where  
$\vec{\eta} = \{\beta_x, \beta_p, \zeta, \kappa\}$ is the set of probe parameters which are accessible in practice and can be adjusted to optimise the estimation precision. 

Let us begin with the simple case of position- or momentum-squeezed probe states, corresponding to the parameters $\{\xi,\kappa\} = \{\zeta,0\}$ or $\{\zeta,\pi\}$, respectively. As we focus on estimating the squeezing amplitude, $r$, the encoding squeezing operator is given by $\hat{S}(z) \equiv \hat{S}(r,0)$.
%
After passing through the unitary channel, the encoded state is given by $|{\phi(\vec{\theta}; \vec{\eta})}\rangle_{DS}
= \hat{D}(\alpha) \hat{S}(r)\hat{D}(\beta) \hat{S}(\xi)\ket{0}$,
or $|{\phi(\vec{\theta}; \vec{\eta})}\rangle_{SD}$
if the order of operations $\hat{D}(\beta)$ and $\hat{S}(\xi)$ are interchanged. 
The respective SLD-CRBs read
\begin{eqnarray}
    C^{P}_{DS}(\vec{\theta}; \vec{\eta}) &=& \cosh^2 (r \pm \zeta) + \frac{e^{2r}\beta_x^2 + e^{-2r}\beta_p^2}{2}~~\mathrm{and}
    \label{eq:Gaussian_sld} \\
    C^{P}_{SD}(\vec{\theta};\vec{\eta})&=& \cosh^2 (r \pm \zeta) + \frac{e^{2(r \pm \zeta)}\beta_x^2 + e^{-2(r \pm \zeta)}\beta_p^2}{2}, \nonumber 
    \label{eq:Gaussian_sld_reverse_order} \\
\end{eqnarray}
where $\pm \zeta$ corresponds to position and momentum squeezing in the probe preparation, respectively (see Appendix~\ref{app:SLD_Gaussian} for the full derivation). Here, the superscript $P$ indicates that we are considering the SLD-CRB for different probe configurations, while the encoding is fixed to $\hat{D}(\alpha) \hat{S}(r)$. The subscripts specify the order of Gaussian operations during probe preparation. 

A few immediate observations emerge from the above expressions. Firstly, both SLD-CRBs are independent of the encoding displacement parameter $\alpha$ due to symmetry~\cite {Bressanini_JPA_2024_Gaussian_multiparameter}.  
Secondly, the expressions of the SLD-CRBs clearly show that for pure Gaussian probes used to estimate displacement and squeezing, the SLD-CRB corresponding to momentum squeezed states is always lower than that of the position squeezed states, since $C^{P}_{DS}(\vec{\theta}; \vec{\eta} = \{\vec{\alpha}, \zeta\}) - C^{P}_{DS}(\vec{\theta}; \vec{\eta} = \{ \vec{\alpha}, -\zeta \}) = \sinh 2r \sinh 2 \zeta \geq 0$, and similarly for $C^{P}_{SD}$. 
In fact, the probe squeezing strength, $\zeta$, can have a constructive impact on the sensing protocol, unlike the encoding squeezing parameter, $r$, which always increases the SLD-CRBs by reducing the variance of one quadrature at the expense of the other~\cite{Bressanini_JPA_2024_Gaussian_multiparameter}. 
Specifically, the minimum SLD-CRB is attained by momentum-squeezed vacuum states with the same squeezing strength as the encoding unitary, i.e., $\zeta = r$ and $\beta_x = \beta_p = 0$, from where we obtain $C^{P}_\mathcal{G}(\vec{\theta}; -\zeta)_{\min} = 1$. In fact, no other probe squeezing angle different from $\kappa = \pi$ can furnish a lower error bound. One can thus tune the probe squeezing to an optimal value to nullify the destructive effect of the encoding squeezing and reduce the SLD-CRBs.  This, however, cannot constitute a legitimate step in the protocol since the encoding squeezing parameter is unknown a priori. However, it has been shown that one can initially estimate the value of the squeezing parameter through sequential measurements~\cite{Rodriguez_Quantum_2024} such that the proper squeezing of the probe state can be chosen to attain the minimum SLD-CRB. 





Finally, comparing the SLD-CRBs of the two different probe preparation strategies, it can be observed that the probe squeezing strength, $\zeta$, determines which SLD-CRB is lower, i.e, which configuration of the probe state, Eq.~\eqref{eq:Gaussian_sld} or Eq.~\eqref{eq:Gaussian_sld_reverse_order}, is better for the estimation protocol. It is straightforward to show that, for a given probe-squeezing direction, $\ket{\phi (\vec{\eta})}_{SD}$ can outperform $\ket{\phi (\vec{\eta})}_{DS}$, under the following conditions
\begin{eqnarray}
C^{P}_{SD} \Big( \vec{\theta}; \vec{\eta} = \{ \beta,  \zeta\} \Big) \leq C^{P}_{DS} \Big( \vec{\theta}; \vec{\eta} = \{ \beta,  \zeta\} \Big), 
   \label{eq:SD_less-than_DS_x}
\end{eqnarray}
when $\zeta \leq \log \Big| \frac{e^{-2r} \beta_p}{\beta_x} \Big|$ {for squeezing along} $\hat{x}$,
and similarly,
%
\begin{eqnarray}
C^{P}_{SD} \Big( \vec{\theta}; \vec{\eta} = \{ \beta,  -\zeta\} \Big) \leq C^{P}_{DS} \Big( \vec{\theta}; \vec{\eta} = \{ \beta,  -\zeta\} \Big), 
   \label{eq:SD_less-than_DS_p}
\end{eqnarray}
{when} $\zeta \geq \log \Big| \frac{e^{2r} \beta_x}{\beta_p} \Big|$ {for squeezing along} $\hat{p}$.
However, 
ignorance of $r$ forbids predetermining the probe configuration to attain better precision. 

\subsection{Optimal pure Gaussian probes}
\label{subsec:op_optimal}

Let us investigate the optimal Gaussian probe state that minimizes both the SLD-CRBs in Eqs.~\eqref{eq:Gaussian_sld} and~\eqref{eq:Gaussian_sld_reverse_order}.
Identifying the parameter regimes that furnish the least $C^{P}_{DS}(\vec{\theta}; \vec{\eta})$ and $C^{P}_{SD}(\vec{\theta}; \vec{\eta})$ requires keeping the corresponding probe at a fixed energy, since in CV systems, arbitrary displacement and squeezing can produce states with unbounded energy, which is not physical. For brevity, we will drop the arguments in the SLD-CRB expression and only write $C^{P}_{DS} (C^{P}_{SD})$, unless explicitly required.

The energy of the Gaussian probe states are $\mathcal{E}^{P}_{DS} = \sinh^2{\zeta} + \beta_x^2 + \beta_p^2$ and $\mathcal{E}^{P}_{SD} = \sinh^2{\zeta} + e^{\pm \zeta} \beta_x^2 + e^{\mp \zeta} \beta_p^2$. Since the encoding parameters are not known, the energy of the encoding is not considered. For the squeezed-coherent state, the direction of squeezing determines the energy of the state. In contrast, for the displaced-squeezed state, the energy is independent of the squeezing direction. In Appendix~\ref{app:optimal}, we tackle the minimization of $C^{P}_{DS}(C^{P}_{SD})$, with respect to the probe parameters, $\vec{\eta} = \{\beta_x, \beta_p, \pm \zeta \}$, at a fixed $\mathcal{E}^{P}_{DS} (\mathcal{E}^{P}_{SD}) = \mathcal{E}^{P}$. \\

\textbf{Observation.} {\it For estimating displacement and squeezing simultaneously, the optimal single-mode pure Gaussian probe with energy, $\mathcal{E}^P$, involves both squeezing and displacement operations along the momentum quadrature, i.e, $\ket{\phi(\vec{\eta})}_{\mathcal{G}}^{\mathrm{opt}}$ is either $\ket{\phi(\vec{\eta})}^{\mathrm{opt}}_{DS} = \hat{D}(\iota \beta_p) \hat{S}(-\zeta_\mathrm{opt}) \ket{0}$ or $\ket{\phi(\vec{\eta})}^{\mathrm{opt}}_{SD} = \hat{S}(-\zeta_\mathrm{opt}) \hat{D}(\iota \beta'_p) \ket{0}$}.\\

The SLD-CRB for this choice of optimal single-mode pure energy-constrained Gaussian states, with squeezing strength $\zeta_\mathrm{opt} = \log (2 e^{4r} - 1)/4$ along the momentum quadrature, reads
\begin{equation}
    C^{P}_{\mathcal{G}, \mathrm{opt}} = \frac{1}{2} \Bigg( 1 + \frac{e^{-2r}}{2} (2 \mathcal{E}^P + 1) + \frac{\sqrt{2 - e^{-4r}}}{2} \Bigg),
    \label{eq:SLD-CRB_opt}
\end{equation}
while the total energy constraint determines the respective displacement parameters.


\textbf{Remark $\mathbf{1}$.} It is quite intriguing that, no matter which probe configuration we start with, at a fixed energy, the optimal SLD-CRB for a generic pure Gaussian state is unique. While interchanging the order of displacement and squeezing operations essentially creates the same state, but with a change in the parameter values,
the energy of a displaced-squeezed state and that of a squeezed-coherent state are not exactly the same. Therefore, having a unique optimal SLD-CRB for pure Gaussian states reveals a beautiful symmetry in the setup, a fact that can be studied in future works, especially from the point of view of identifying quantum resources which dictate the precision of the protocol.

\begin{figure}[t]
\includegraphics[width=.9\linewidth]{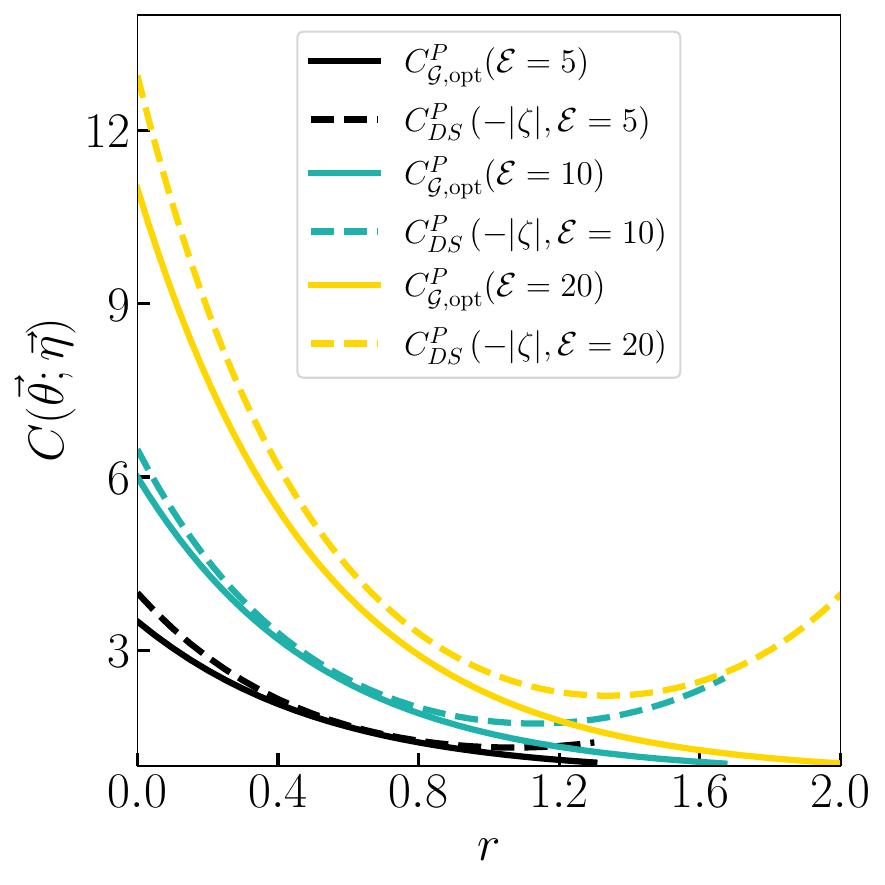}
\captionsetup{justification=Justified,singlelinecheck=false}
\caption{\textbf{SLD-CRB for single-mode pure Gaussian states employed in multiparameter estimation.} The SLD-CRB, $C(\vec{\theta}; \vec{\eta})$ (ordinate), is shown against the encoding squeezing strength, $r$ (abscissa), for the optimal Gaussian probe state, $C^{P}_{\mathcal{G}, \mathrm{opt}}$ (solid line), and the displaced-squeezed state with momentum squeezing,  $C_{DS}(-\zeta)$ (dashed line). We consider probes of different energy, $\mathcal{E}^P = 5$ (dark black), $\mathcal{E}^P = 10$ (light blue),  and $\mathcal{E}^P = 20$ (very light yellow). The displaced-squeezed probe parameters for the different energies are given as $\{ \beta_x = 0.1,~ \beta_p = 2 \}_{\mathcal{E}^P = 5}$, $\{ \beta_x = 0.2, ~\beta_p = 3 \}_{\mathcal{E}^P = 10}$, and $\{ \beta_x = 0.3,~ \beta_p = 4 \}_{\mathcal{E}^P = 20}$, where in each case, the momentum-squeezing strength is calculated according to $\zeta = \sinh^{-1} \sqrt{\mathcal{E}^P - |\beta|^2 }$. Both axes are dimensionless.}
\label{fig:opt_Gaussian}
\end{figure}

For any single-mode pure Gaussian probe, $C^{P}_{\mathcal{G}}(\vec{\theta}; -\zeta)_{\min} = 1$, which must also be satisfied by Eq.~\eqref{eq:SLD-CRB_opt}. For a particular value of $r$,  this
imposes a lower bound on the energy of the probe state, such that
\begin{equation}
    \mathcal{E}^P \geq \frac{1}{2} \Bigg( e^{2r }(2 - \sqrt{2 -  e^{-4r}}) - 1 \Bigg),
    \label{eq:energy_condition}
\end{equation}
which
yields $C^{P}_{\mathcal{G}, \mathrm{opt}} = 1$. Further, Eq.~\eqref{eq:energy_condition} 
implies $\mathcal{E}^P \geq \sinh^2 \zeta_\mathrm{opt}$, which is satisfied by the optimal displaced-squeezed or squeezed-coherent probe state.
However, 
a probe state $|\phi(\beta_p, -\zeta_\mathrm{opt}^{-})\rangle_{DS (SD)}$ of energy $\mathcal{E}^P$ can only estimate up to a particular maximum value of the encoding squeezing strength, given by
\begin{equation}
    r \leq r_{\max} = \frac{1}{2} \log \Bigg( 1 + 2 \mathcal{E}^P + \sqrt{2 \mathcal{E}^P (1 + \mathcal{E}^P)} \Bigg).
    \label{eq:r_condition}
\end{equation}
This implies that higher energy probes can estimate larger values of the squeezing strength $r$, which is illustrated in Fig.~\ref{fig:opt_Gaussian}. We observe that for lower energies, the SLD-CRB is computable only up to a threshold value,
$r_{\max}(\mathcal{E}^P = 5) \approx 1.32,~ r_{\max}(\mathcal{E}^P = 10) \approx 1.67, ~\text{and}~ r_{\max}(\mathcal{E}^P = 20) \approx 2.01$.

\subsection{Energy dependence and operational consequences of the optimal probe setup}
\label{subsubsec:energy_optimal}

The optimal Gaussian SLD-CRB is a monotonically increasing function of the pre-encoding energy, $\mathcal{E}^P$, of the probe (see Fig.~\ref{fig:opt_Gaussian}). The immediate question then is whether vacuum can be the most beneficial probe since it carries the least energy. This is not true, however. The SLD-CRB for vacuum, $C^P_{\ket{0}} = \cosh^2 r$, can be beaten by that of the momentum-squeezed state, $C^P_{S(-\zeta)} = \cosh^2 (r - \zeta)$; in fact, while $C^P_{\ket{0}}$ increases with $r$, the momentum-squeezed probe can outperform vacuum over a greater range of squeezing $\zeta \in (0, 2r)$, which, in turn, implies that the energy, $\sinh^2 \zeta$, does have a constructive impact on the protocol. A similar benefit of high-energy probes has also been discussed above, where we show that a higher energy 
implies that a larger range of $r$ can be estimated. Furthermore, the gap between $C^P_{\mathcal{G}, \mathrm{opt}}$ and $C^{P}_{DS}$ increases with the increase in energy (see Fig.~\ref{fig:opt_Gaussian}), highlighting another advantage of using high-energy probes whose optimal accuracy is far superior to that of the displaced-momentum-squeezed probe. Finally, it can be easily shown that $C^{P}_{\mathcal{G}, \mathrm{opt}} \leq C^P_{S(-\zeta)} ~\forall~ \mathcal{E}^P$ when Eq.~\eqref{eq:energy_condition} is respected. This also has a convenient operational consequence in that, for a given energy, $\mathcal{E}^P$, the most precise probe is one in which the energy is distributed between squeezing and displacement operations. Since squeezing is experimentally costly, this division of energy makes the optimal probe the most advantageous one, not only in terms of precision, but also in terms of practical considerations.


\textbf{Note $\mathbf{1}$.} A quick look at Eq.~\eqref{eq:SLD-CRB_opt} suggests that the optimal SLD-CRB, $C^P_{\mathcal{G}, \mathrm{opt}}$, for an energy-constrained single-mode Gaussian state $\ket{\phi(\vec{\eta})}^{\mathrm{opt}}_{DS (SD)}$ decreases with $r$. This seems counterintuitive since it implies that the estimation precision increases with an increase in the parameter to be estimated. We have already discussed how more energetic probes are necessary to estimate higher values of the encoding squeezing strength, and an immediate expectation is that high $r$ will also cause the estimation protocol to become less effective. However, the decrease of the optimal SLD-CRB with $r$ can be explained in terms of the quantum Fisher information. The QFI is an indication of how fast a quantum state changes when its parameters are changed by a small amount. At low $r$, the phase-space representation of the state is not much deformed from its original form, and a small change, say $\delta r$, does not alter the deformation appreciably. Therefore, the states with encoding squeezing strength $r$ and $r + \delta r$ are close to each other in the Hilbert space, which makes distinguishing them a difficult task. On the other hand, for large $r$, the probe is extremely disturbed from its initial phase-space representation, and a tiny change, $\delta r$, causes the deformation to change drastically. Thus, at high encoding squeezing strength, the states with $r$ and $r + \delta r$ encodings are far apart from each other in the Hilbert space, which makes distinguishing them easier. This causes the Fisher information to increase rapidly with changes in $r$, whence the SLD-CRB decreases, and the estimation becomes more accurate.

\section{Multiparameter estimation through indefinite causal order in the probe preparation setup}
\label{sec:ICO}

In the preceding sections, we have derived the optimal Gaussian probe for sensing squeezing and displacement parameters, demonstrating that the order in which the probe-preparation unitaries were applied determined the eventual precision. Recall that the encoding operation was fixed to $\Lambda(\vec{\theta}) = \mathcal{D}(\alpha) \circ \mathcal{S}(r)$.
Quantum mechanics, however, allows us to incorporate superposition in the order of processes comprising sequential operations, known as {\it indefinite causal order} (ICO)~\cite{Oreshkov_NC_2012_correlations_ICO, Colnaghi_PLA_2012_computation_IC0_gates, Chiribella_PRA_2013_computation_ICO, Baumeler_IEEE_2014_signalling_ICO, Oreshkov_NJP_2016_causal_separable, Bisio_PRSA_2019_higher-order_quantum-theory}. 
Although motivated by the aim to unify general relativity and quantum theory~\cite{Hardy_JPA_2007_quantum-gravity_ICO, Hardy_book_2009_quantum-gravity_ICO}, ICO has been shown to offer advantage over information-theoretic protocols with a definite causal order in communication~\cite{Guerin_PRL_2016_switch_communication, Ebler_PRL_2018_ICO_communication, Salek_arXiv_2018_ICO_communication, Chiribella_PRSA_2019_ICO_Shannon, Goswami_PRR_2020_ICO_communication, Chiribella_NJP_2021_ICO_communication}, channel discrimination~\cite{Chiribella_PRA_2012_ICO_channel-discrimination, Arajo_PRL_2014_ICO_channel}, non-local games~\cite{Gurin_NJP_2018_ICO_non-locality, Zych_Nature_2019_ICO_Bell-theorem}, and even in designing quantum thermal machines~\cite{Felce_PRL_2020_ICO_refrigeration, Guha_PRA_2020_ICO_thermal, Simonov_PRA_2022_ICO_work-extraction, Francica_PRA_2022_ICO_work-extraction, Zhu_PRL_2023_ICO_charging, Simonov_NJP_2025_ICO_thermal}. 

In the study of quantum metrology, ICO has been theorised to provide super-Heisenberg scaling for sensing multiple displacements~\cite{Zhao_PRL_2020_ICO_metrology}, which has also been shown in experiments~\cite{Yin_2023}. One of the principal advantages offered by ICO in metrology is against noise~\cite{Chapeau-Blondeau_PRA_2021_ICO_noisy-metrology, Chapeau-Blondeau_PRA_2021_ICO_noisy-metrology2, Chapeau-Blondeau_PLA_2022_ICO_noisy-metrology3, Chiribella_arXiv_2022_ICO_noisy-metrology, Goldberg_PRR_2023_ICO_noisy-metrology, Kurdzialek_PRL_2023_ICO_noisy-metrology}, for which an indefinite ordering among channels can help to achieve a better precision~\cite{Liu_PRL_2023_optimal_metrology_strategy}. Furthermore, ICO, in the form of time-flip~\cite{Chiribella_CP_2022_indefinite_time-direction, Liu_NJP_2023_indefinite_time-direction_communication}, has also been applied to attain Heisenberg-limited precision in single-parameter estimation~\cite{Agrawal_Quantum_2025_time-directed_metrology}. Despite such rapid progress, however, there are instances where the claim for advantage through ICO has been under scrutiny~\cite{Mukhopadhyay_arXiv_2018_ICO_metrology, Frey_QIP_2021_ICO_noisy-metrology}, and the metrological advancements leverageable from different kinds of processes have been delineated~\cite{Mothe_PRA_2024_ICO_metrology_processes}.

Here, we investigate whether ICO can provide any advantage over conventional protocols with a fixed order of operations in the considered multiparameter estimation protocol. 
The indefiniteness can be implemented in the ordering of the displacement and squeezing unitaries, during either the $(1)$ probe preparation or $(2)$ the encoding operations. 
This can be done applying a
quantum switch~\cite{Chiribella_PRA_2013_computation_ICO, Procopio_NC_2015_experimental_switch, Guerin_PRL_2016_switch_communication, Wei_PRL_2019_experiment_switch_communication}, where an auxiliary quantum system is employed to coherently control the temporal ordering of the constituent operations. Specifically, given the displacement, $\mathcal{D}$, and squeezing channels, $\mathcal{S}$, channels, the quantum switch is prepared by coupling the intended system, say $\ket{\psi}$, to a qubit state, $\ket{s} = \sqrt{p} \ket{0} + \sqrt{1 - p} \ket{1}$ (with $0 \leq p \leq 1$), and subjecting the joint system to the ICO channel given by
\begin{eqnarray}
    \Lambda_{\mathrm{ICO}}:= \mathcal{D} \circ \mathcal{S} \otimes \ketbra{0}{0} + \mathcal{S} \circ \mathcal{D} \otimes \ketbra{1}{1}
    \label{eq:ICO_defn}
\end{eqnarray}
that acts on $\ket{\psi} \otimes \ket{s}$. Finally, the qubit is measured in the basis $\{\ket{\pm} = (\ket{0} \pm \ket{1})/\sqrt{2} \}$, yielding the unnormalized output states
\begin{eqnarray}
    \ket{\psi_\pm} \sim \sqrt{\frac{p}{2}} (\mathcal{D} \circ \mathcal{S}) [\ket{\psi}] \pm \sqrt{\frac{1 - p}{2}} (\mathcal{S} \circ \mathcal{D}) [\ket{\psi}],
    \label{eq:ICO_state_ex}
\end{eqnarray}
conditioned on the measurement outcomes $\ketbra{+}{+}$ and $\ketbra{-}{-}$ respectively.
Evidently, the ICO parameter, $p$, of the auxiliary qubit determines the sequence of operations as well as the strength of a given order of operations in the output superposition in Eq.~\eqref{eq:ICO_state_ex}. A higher value of $p$ means that the ordering $\mathcal{D} \circ \mathcal{S}$ has a greater weight than its counterpart. In fact, for $p = 0$ and $p = 1$, one recovers the definite orderings, $\mathcal{S} \circ \mathcal{D}$ and $\mathcal{D} \circ \mathcal{S}$ respectively, between the two operations. In the following, we examine how the SLD-CRB behaves with the variation in $p$ when ICO is implemented during probe preparation, and whether ICO can provide any advantage over conventional probe-preparation schemes. 

To investigate the impact of incorporating ICO during probe preparation, we keep the encoding order of operations fixed to $\Lambda(\vec{\theta}) = \mathcal{D} \circ \mathcal{S}$, which, however, is applied on $\ket{\phi(\vec{\eta})}_{\pm}$, the probe state prepared using the quantum switch
\begin{eqnarray}
   \ket{\phi(\vec{\eta})}_{\pm} \sim \Big(\sqrt{\frac{p}{2}} \hat{D}(\beta) \hat{S}(\xi) \pm \sqrt{\frac{1 - p}{2}} \hat{S}(\xi) \hat{D}(\beta)\Big) \ket{0},\nonumber\\
   \label{eq:ICO_probe}
\end{eqnarray}
where $\xi \in \{ \pm \zeta \}$ as before, with its positive and negative values indicating position and momentum squeezing, respectively, and $\ket{\phi(\vec{\eta})}_{+}$ and $\ket{\phi(\vec{\eta})}_{-}$, represent states comprising even and odd parity superpositions, respectively. The normalization factors are given as
\begin{eqnarray}
   \nonumber N_\pm &=& \frac{1}{2} \pm \sqrt{p(1 - p)} \cos \Big( 2 \beta_x  \beta_p \sinh \xi \Big) \\
   &+& \nonumber \exp \Big( - \frac{1}{2} e^{2\xi} (e^{\xi} - 1)^2 \times \\
   && ~~~~~~~(\beta_x^2 + e^{2 \xi} \beta_p^2) \Big). 
   \label{eq:ICO_probe_normalisation}
\end{eqnarray}
The SLD-CRB, $C^{P_\pm}_{\mathrm{ICO}}$, is too complicated to present here, but we provide the expressions for the SLDs in Appendix~\ref{app:SLD_ICO}. 

\subsection {Enhanced precision through indefinite causal order in probe preparation}
\label{subsec:ICO_probe}

Let us first analyze how the SLD-CRBs corresponding to $\ket{\phi(\vec{\eta})}_{\pm}$ vary with respect to the parameter $p$, which determines the weight of a particular operation sequence in the probe prepared through the quantum switch. For this purpose, we consider a fixed set of encoding parameters as $\alpha_x = 0.5$, $ \alpha_p = 0.4$, $ r = 0.2$.  Note that for higher values of the encoded parameters, the SLD-CRBs assume higher values, and vice versa for lower parameter values. However, the qualitative features of the SLD-CRBs remain unchanged from those reported hereafter. We shall first analyze the SLD-CRBs for the ICO probe with even superposition, $\ket{\phi(\vec{\eta})}_+$, and with odd superposition, $\ket{\phi(\vec{\eta})}_-$, before comparing the scheme with the optimal pure Gaussian probe. For this purpose, we compute $C^{P_\pm}_{\mathrm{ICO}}$ numerically, in a truncated Hilbert space spanned by the first $\{\ket{n}\}_{n = 0}^{100}$ Fock states. We claim that ICO provides an {\it advantage} over the definite-ordered protocol when $C^{P_\pm}_{\mathrm{ICO}} \leq \{C^{P}_\text{DS},~ C^P_{SD}\}$, i.e., the SLD-CRB for the ICO setup is lower than the SLD-CRBs of {\it both} the Gaussian probe states prepared with a fixed order of operations.

\begin{figure*}[t]
\includegraphics[width=0.6\linewidth]{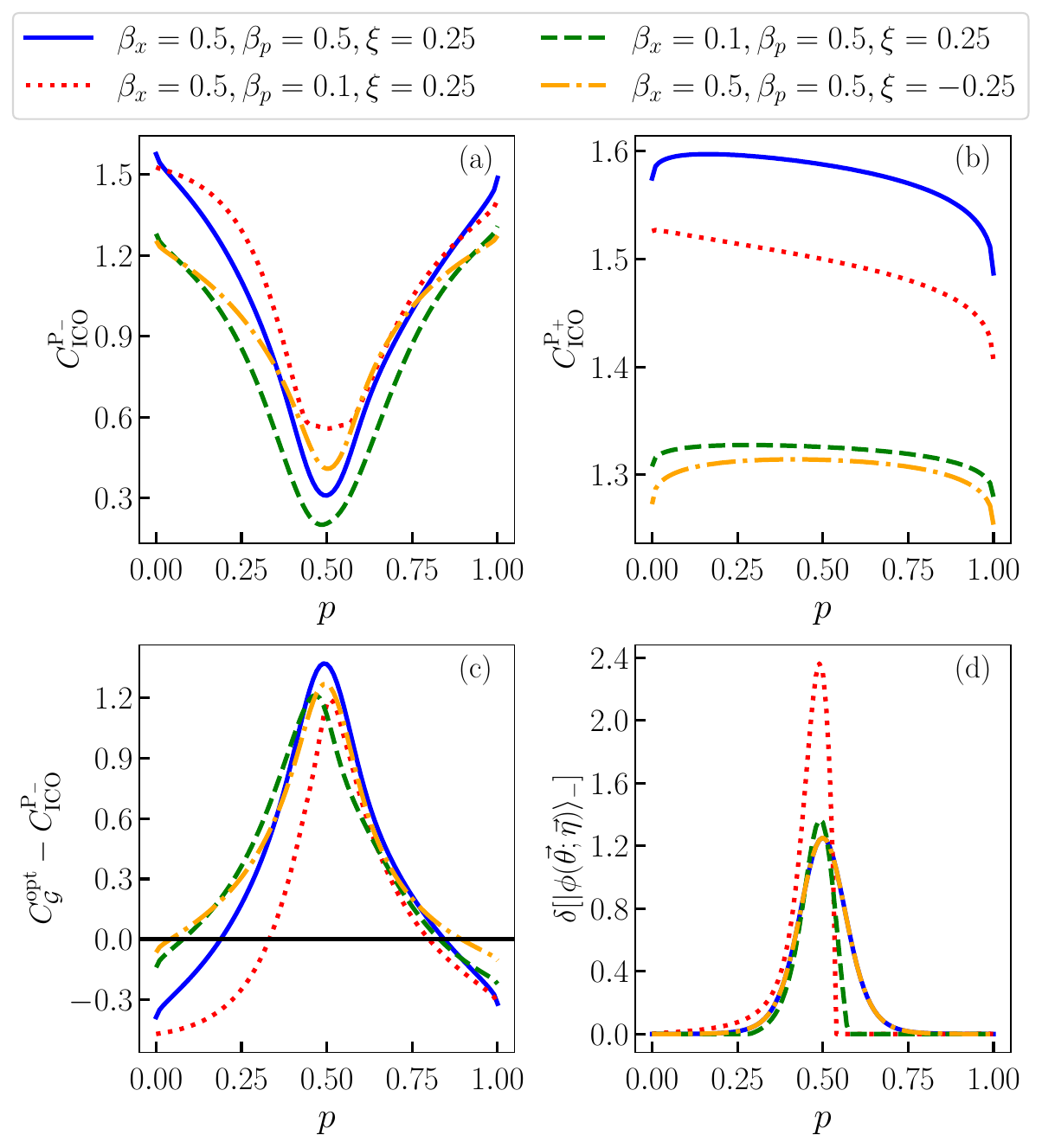}
\captionsetup{justification=Justified,singlelinecheck=false}
\caption{\textbf{SLD-CRB for probes prepared through ICO.} (a) $C^{P_-}_{\mathrm{ICO}}$ (ordinate) is shown against the ICO parameter, $p$ (abscissa), for the odd superposition ICO probe. (b) $C^{P_+}_{\mathrm{ICO}}$ (ordinate) with respect to $p$ (abscissa), for the even superposition. (c) The SLD-CRB difference, $C^{\mathrm{opt}}_{\mathcal{G}} - C^{P_-}_{\mathrm{ICO}}$(ordinate), is shown against $p$ (abscissa), between the odd superposition ICO probe and the optimal pure displaced-squeezed Gaussian probe. For reference, $0$ is depicted as a black solid line to indicate where the difference vanishes. (d) The relative entropy of non-Gaussianity, $\delta[|\phi(\vec{\theta}; \vec{\eta})\rangle_-]$ (ordinate), for the odd superposition ICO probe is plotted against $p$ (abscissa). In all the subplots, the probe parameters are given by $\{\beta_x = 0.5,~ \beta_p = 0.5,~ \xi = 0.25\}$ (blue solid line), $\{\beta_x = 0.5,~ \beta_p = 0.1,~ \xi = 0.25\}$ (red dotted line), $\{\beta_x = 0.1,~ \beta_p = 0.5,~ \xi = 0.25\}$ (green dashed line), and $\{\beta_x = 0.5,~ \beta_p = 0.5,~ \xi = -0.25\}$ (orange dash-dotted line). The encoding parameters have been fixed to $\{\alpha_x = 0.4, ~\alpha_p = 0.5,~ r = 0.2\}$ throughout. Both axes are dimensionless. }
\label{fig:ICO_probe}
\end{figure*}

{\bf $\mathbf{1.}$ For $\ket{\phi(\vec{\eta})}_-$:} The odd superposition from the switch-based probe preparation scheme yields ubiquitous advantage over both definite-ordered Gaussian probes. (see Fig.~\ref{fig:ICO_probe}(a)).  Specifically, for all probe configurations, the ICO probe can always outperform both $\ket{\phi(\vec{\eta})}_{SD}$ and $\ket{\phi(\vec{\eta})}_{DS}$, i.e., $\forall~ p, C_{\mathrm{ICO}}^{P_-} \leq \{C^P_{DS},~ C^P_{SD}\}$. The most prominent advantage of this ICO probe configuration is the decrease of the SLD-CRB below $1$, which is the minimum amount of error achievable by any Gaussian probe state. Thus, the ICO shows a constructive impact on the estimation protocol when the output state comprises the odd superposition. 

In this situation, the configuration with more position displacement, $\{\beta_x > \beta_p \}$, performs the worst, but still outperforms the pure Gaussian probes. The odd superposition ICO probe with position squeezing and more displacement along the momentum quadrature, $\{ \beta_x < \beta_p, ~ \xi \geq 0\}$, furnishes the lowest SLD-CRB, while the precision offered by the probe with equal displacement along both quadratures is intermediate to the two aforementioned cases. Notice that, akin to the optimal pure Gaussian probe, squeezing along the momentum quadrature allows us to attain better precision. Further, increasing $\beta_x$ causes the SLD-CRB to increase (comparing $\beta_x = 0.1$ and $\beta_x = 0.5$ configurations). 

{\bf $\mathbf{2.}$ For $\ket{\phi(\vec{\eta})}_+$:} Comparing the SLD-CRBs of $\ket{\phi(\vec{\eta})}_+$ for different probe-preparation configurations, we observe that when the squeezing is along the momentum quadrature, i.e., $\{ \xi < 0 \}$, the considered probe state works best (see Fig.~\ref{fig:ICO_probe}(b)). For $0 < p < 1$, the SLD-CRB is always lower than $C^P_{SD}$ for the chosen set of probe and encoding parameters, which implies that ICO can outperform the squeezed-coherent probe state. However, the probe can never outperform $\ket{\phi(\vec{\eta})}_{DS}$ (at $p = 1$). In fact, since the parameters do not adhere to Eqs.~\eqref{eq:SD_less-than_DS_x} and~\eqref{eq:SD_less-than_DS_p}, we always have $C^P_{SD} > C^P_{DS}$, and surpassing the accuracy offered by $\ket{\phi(\vec{\eta})}_{SD}$ does not guarantee the same over $\ket{\phi(\vec{\eta})}_{DS}$. For all configurations, ICO has a detrimental effect on the estimation precision because the SLD-CRB is minimum at the limits $p = 0$ and $p = 1$, which depict probes prepared with a definite sequence of operations. 

Contrary to the optimal pure Gaussian probe, displacement and squeezing along the momentum quadrature do not aid in enhancing the accuracy of the protocol. Moreover, the value of $C^{P_+}_{\mathrm{ICO}}$ is an order of magnitude higher than that of the odd superposition ICO probe, which immediately implies that $\ket{\phi(\vec{\eta})}_-$ can estimate the parameters with a much better precision than $\ket{\phi(\vec{\eta})}_+$ (comparing Figs.~\ref{fig:ICO_probe}(a) and (b)). Therefore, we can conclude that for the ICO probe, comprising an even superposition between the two conventional schemes, there is no advantage to introducing indefiniteness in the probe-preparation strategy. If the measurement outcome during the quantum switch protocol happens to be $\ketbra{+}{+}$, it is best to employ $\ket{\phi(\vec{\eta})}_{DS}$ instead of the output state.


\textbf{Remark $\mathbf{2}$.} It should be noted that, the energy of $\ket{\phi(\vec{\eta})}_{\pm}$ prepared through ICO is different from that of $\ket{\phi(\vec{\eta})}_{DS(SD)}$. Specifically, the even superposition has energy intermediate between the displaced-squeezed and squeezed coherent Gaussian probes. On the other hand, the odd superposition ICO probe has energy lower than both the pure Gaussian probes. Thus, while the energy definitely plays a part in determining the precision offered by a probe, it is not the only parameter that guarantees a lower SLD-CRB. Other quantum mechanical properties should also be taken into consideration when trying to explain the resource behind the advantage in the multiparameter estimation protocol.

{\bf $\mathbf{3.}$ Comparison with $\ket{\phi(\vec{\eta})}^{\mathrm{opt}}_{\mathcal{G}}$:} Since we can argue that the ICO probe with odd superposition can outperform both the definite-ordered Gaussian probes, as well as that prepared through even superposition, we compare the optimal pure Gaussian probe with $\ket{\phi(\vec{\eta})}_-$. For a fair comparison, we must ensure that both the probes under study have the same energy, since the optimality is determined with that constraint in Sec.~\ref{sec:disp_sq_Gaussian}. To do so, we estimate the energy of the odd superposition for every value of $p$, and at each such $p$, compute the optimal SLD-CRB, $C^{P}_{\mathcal{G}, \mathrm{opt}}$, with $\mathcal{E}^P = \mathcal{E}(\ket{\phi(\vec{\eta})}_-)$ in Eq.~\eqref{eq:SLD-CRB_opt}. 

We analyze the difference, $C^{P}_{\mathcal{G}, \mathrm{opt}} - C^{P_-}_{\mathrm{ICO}}$ (see Fig.~\ref{fig:ICO_probe}(c)), wherein a positive value of the same indicates ICO advantage. We observe that when $\beta_p > \beta_x$, the odd superposition ICO probe furnishes a lower SLD-CRB than the optimal Gaussian one when $0.2 \lesssim p \lesssim 0.8$, with the advantage being prominent around $p = 0.5$. Therefore, not only can ICO outperform the Gaussian probes with a definite order of operations, but they can also estimate with greater precision than the optimal pure Gaussian probe at the same energy. For simplicity, we have refrained from illustrating the SLD-CRBs for the even superposition case, as well as for the odd one with $\beta_p < \beta_x$, since they can never outperform the optimal Gaussian probe. 

\subsection{Does non-Gaussianity act as a resource?}
\label{subsec:ICO_probe_nG}

The quantum switch setup comprises Gaussian unitaries and vacuum as the initial state, but its output is definitely non-Gaussian, being the coherent superposition of two Gaussian states. Non-Gaussian states~\cite{Walschaers_PRX_2021_non-Gaussian}, such as Fock states~\cite{Rahman_PRL_2025_metrology_Fock}, cubic-phase states~\cite{Guo_arXiv_2025_metrology_cubic-phase-state}, and Gottesman-Kitaev-Preskill (GKP) states~\cite{Labarca_PRXQ_2026_GKP_metrology}, have been shown to offer higher precision as compared to Gaussian states in metrological setups~\cite{Park_arXiv_2025_phase-estimation_non-gaussian, Grochowski_PRL_2025_phase-insensitive_non-Gaussian_metrology}. This prompts us to ask whether the gain through ICO stems from the non-Gaussianity present in the output probe state. 

To address this question, we choose the relative entropy of non-Gaussianity~\cite{Genoni_PRA_2008_relative-entropy} to investigate this aspect, which is a monotone under Gaussian transformations.
It is defined as the minimum distance, in terms of the relative entropy, of the state from its closest Gaussian counterpart. Given a non-Gaussian state, $\rho$, its reference Gaussian state, $\rho_{\mathcal{G}}$, is the one with the same displacement vector and covariance matrix, i.e., $\vec{d}(\rho_{\mathcal{G}}) = \vec{d}(\rho)$ and $\Xi(\rho_{\mathcal{G}}) = \Xi(\rho)$. The relative entropy of non-Gaussianity is then the relative entropy between $\rho$ and $\rho_\mathcal{G}$ given as~\cite{Genoni_PRA_2008_relative-entropy}
\begin{eqnarray}
    \delta[\rho] = S(\rho || \rho_{\mathcal{G}}) = S(\rho_{\mathcal{G}}) - S(\rho),
    \label{eq:rel-ent_nG}
\end{eqnarray}
where $S(\sigma) := -\Tr[ \sigma \log_2 \sigma]$ is the von-Neumann entropy~\cite{nielsen_2010} of a density matrix, $\sigma$, and $S(\sigma_1 || \sigma_2) := \Tr [\sigma_1 (\log_2 \sigma_1 - \log_2 \sigma_2)]$~\cite{Vedral_RMP_2002_relative-entropy} is the relative entropy. The second equality follows from the fact that $\rho_\mathcal{G}$ has the same first and second moments as $\rho$~\cite{Holevo_PRA_1999_Gaussian-channel_capacity}. For pure states, it simply reduces to the von-Neumann entropy of the reference Gaussian state~\cite{Genoni_PRA_2008_relative-entropy}. 

Our numerical results for the variation of $\delta[|\phi(\vec{\theta}; \vec{\eta})\rangle_-]$ of the odd-superposition ICO probe state indicate that non-Gaussianity is neither necessary nor sufficient to surpass the optimal Gaussian state in terms of the SLD-CRB. Comparing Figs.~\ref{fig:ICO_probe}(c) and (d), we observe that for probe states with appreciable displacement and squeezing ($\beta_x = \beta_p = 0.5,~ \zeta = 0.25$), $C^{\mathrm{opt}}_{\mathcal{G}} - C^{P_-}_{\mathrm{ICO}} \geq 0 \implies \delta[|\phi(\vec{\theta}; \vec{\eta})\rangle_-] \geq 0$ but not vice-versa, indicating that non-Gaussianity is not a sufficient resource for ICO advantage. On the other hand, for ICO probes with greater displacement along any one direction ($\beta_x = 0.1 < \beta_p = 0.5 ~\text{or vice versa}$), ICO advantage exists even when the relative entropy vanishes, showing that it is not necessary either. For instance, when $\beta_x > \beta_p$, we have $\delta[|\phi(\vec{\theta}; \vec{\eta})\rangle_-] \to 0$ for $p \in (0.58, 0.79)$ where the ICO advantage is still present. Therefore, the resource incorporated by the ICO protocol, which can properly predict advantage, may be a combination of different quantum properties, and cannot be explained by non-Gaussianity alone. As a side note, let us mention that for the even superposition probe, the relative entropy of non-Gaussianity is never non-zero, nor does it provide any advantage over the optimal Gaussian state.

\begin{figure*}[t]
\includegraphics[width=0.6\linewidth]{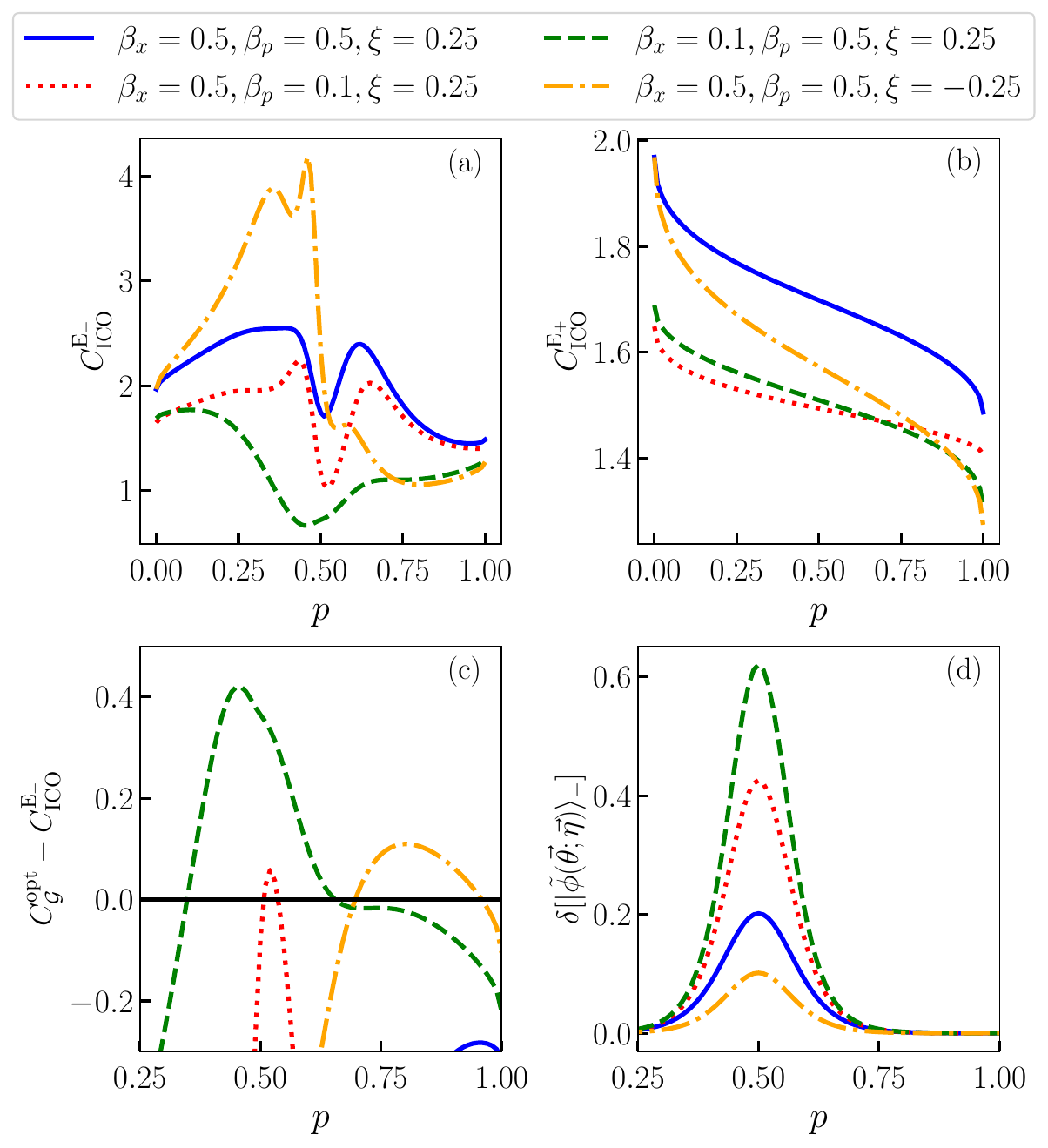}
\captionsetup{justification=Justified,singlelinecheck=false}
\caption{\textbf{SLD-CRB for displaced-squeezed probes after ICO encoding.} (a) The SLD-CRB for the displaced-squeezed probe when considering the odd superposition of ICO encoding, $C^{E_-}_{\mathrm{ICO}}$ (ordinate), is shown against the ICO parameter, $p$ (abscissa). (b) The even superposition encoding case for the same probe, $C^{E_+}_{\mathrm{ICO}}$ (ordinate) with respect to $p$ (abscissa). (c) The SLD-CRB difference between the odd superposition ICO-encoded displaced-squeezed probe and the optimal pure Gaussian probe, $C^{\mathrm{opt}}_{\mathcal{G}} - C^{E_-}_{\mathrm{ICO}}$(ordinate), is shown against $p$ (abscissa). (d) The relative entropy of non-Gaussianity for the odd superposition ICO-encoded probe, $\delta[|\tilde{\phi}(\vec{\theta}; \vec{\eta})\rangle_-]$, is shown against $p$ (abscissa). For reference, $0$ is depicted as a black solid line to indicate where the difference vanishes. All other specifications are the same as in Fig.~\ref{fig:ICO_probe}. Both axes are dimensionless. }
\label{fig:ICO_encoding}
\end{figure*}

\section{Multiparameter estimation with indefinite causal order in the encoding process}
\label{sec:ICO_encoding}

Let us now consider that the encoding process is incorporated into the quantum switch such that the order of encoding operations, which has thus far been fixed to $\mathcal{D}(\alpha) \circ \mathcal{S}(r)$, is made indefinite. As the probe state, we consider the displaced-squeezed vacuum, $\ket{\phi(\vec{\eta})}_{DS} = \hat{D}(\beta) \hat{S}(\xi) \ket{0}$. The channel that encodes the parameters into the probe is given by $\Lambda_{\mathrm{ICO}}(\vec{\theta}):= \mathcal{D}(\alpha) \circ \mathcal{S}(r) \otimes \ketbra{0}{0} + \mathcal{S}(r) \circ \mathcal{D}(\alpha) \otimes \ketbra{1}{1}$, which acts on $\ket{\phi(\vec{\eta})}_{DS} \otimes \ket{s}$, followed by a measurement of the auxiliary qubit $\ket{s} = \sqrt{p} \ket{0} + \sqrt{1 - p} \ket{1}$ in the basis $\{\ketbra{\pm}{\pm}\}$. The unnormalized encoded states corresponding to the two measurement outcomes are 
\begin{eqnarray}
   \nonumber |\tilde{\phi}(\vec{\theta}; \vec{\eta})\rangle_\pm &\sim& \sqrt{\frac{p}{2}} \hat{D}(\alpha) \hat{S}(r) \ket{\phi(\vec{\eta})}_{DS} \\
   &\pm& \sqrt{\frac{1 - p}{2}} \hat{S}(r) \hat{D}(\alpha) \ket{\phi(\vec{\eta})}_{DS}.
   \label{eq:ICO_encoding_state}
\end{eqnarray}
Note that since the encoding itself is done through ICO, the normalisation term of the encoded state is also a function of the parameters to be estimated, i.e., $N_{\pm}(\vec{\theta} = \{ \alpha, r \} )$.
We now numerically estimate the SLD-CRBs corresponding to the even and odd encoding superpositions, $C^{E_\pm}_{\mathrm{ICO}}$,  to investigate whether ICO in the encoding step can furnish any advantage over the fixed orders of encoding (see Appendix~\ref{app:SLD_ICO} for the expressions of the SLDs). Notice that the superscript, $E$, in the SLD-CRB expression indicates that we are now interested in analysing the performance for different encoding operations, while the probe state before encoding has been fixed to $\hat{D}(\beta) \hat{S}(\xi) \ket{0}$ as mentioned beforehand.

We report that an advantage of ICO is present, but not ubiquitous. For the odd superposition encoded state, $|\tilde{\phi}(\vec{\theta}; \vec{\eta})\rangle_-$, initially provides a lower accuracy for $p > 0$ since its SLD-CRB increases with the ICO parameter (see Fig.~\ref{fig:ICO_encoding} (a)). However, around $p \to 0.5$, $C^{E_-}_{\mathrm{ICO}}$ exhibits a minimum, where the precision is better than both the definite-ordered encodings on the displaced-squeezed probe. Therefore, as in the case of ICO in probe preparation, the odd superposition after ICO encoding can provide lower SLD-CRBs as compared to the definite-ordered settings, although such an advantage is available only for a small range of the ICO parameter, contrary to the previous case where the odd superposition ICO-prepared probe was superior for all $p$. The configuration with momentum squeezing is again the most efficient, providing the lowest SLD-CRB around $p \to 0.75$, although appreciable displacement along the momentum quadrature furnishes the low SLD-CRB for the maximum range of $p$.

On the other hand for the even superposition of encoding operations, $|\tilde{\phi}(\vec{\theta}; \vec{\eta})\rangle_+$, we observe that for all configurations of the probe state, the SLD-CRB at any $p > 0$ is always lower than that furnished by the fixed encoding order at $p = 0$, which is, $\hat{S}(r) \hat{D} (\alpha)$ (see Fig.~\ref{fig:ICO_encoding}(b)). 
The lowest $C^{E_+}_{\mathrm{ICO}}$ is apparent over the largest range of the ICO parameter, $0 \leq p \lesssim 0.7$, when the displacement of the initial probe state is more in the position quadrature, contrary to the previous case of $C^{P_+}_{\mathrm{ICO}}$, where more displacement along the momentum quadrature was always more beneficial. However, here too, no configuration can outperform the definite-ordered encoding strategy $\hat{D} (\alpha) \hat{S}(r)$ (at $p = 1$). 

Finally, when we compare $|\tilde{\phi}(\vec{\theta}; \vec{\eta})\rangle_-$ and the optimal pure Gaussian probe, the difference in the corresponding SLD-CRBs peaks around $p \to 0.5$ (see Fig.~\ref{fig:ICO_encoding} (c)), where the ICO-encoded probe has the best precision. Interestingly, although the probe with momentum squeezing furnishes the lowest $C^{E_-}_{\mathrm{ICO}}$, the benefit over the optimal pure Gaussian probe is maximal for states with position squeezing and very low ($\beta_x = 0.1$) displacement along the position quadrature. 

The variation of the relative entropy of non-Gaussianity, $\delta[|\tilde{\phi}(\vec{\theta}; \vec{\eta})\rangle_-]$ (see Fig.~\ref{fig:ICO_encoding} (d)) indicates that non-Gaussianity is again neither necessary nor a sufficient resource for ICO advantage over the optimal Gaussian probe. However, the amount of advantage over the optimal Gaussian probe is proportional to the amount of non-Gaussianity, as evidenced by the SLD-CRB difference for probes with more squeezing along any one quadrature.
This indicates again that there are more quantum properties at play which aid in the ICO protocol, and also, ICO in probe-preparation and encoding constitute two distinct protocols.

\subsection{Role of ICO in mitigating sloppiness}
\label{subsubsec:ICO_sloppiness}

One of the major bottlenecks in precise multiparameter estimation arises when the different parameters are poorly encoded. Specifically, if it is not possible to encode the parameters independently, the resulting correlations act as inherent noise in the protocol, and the extractable information is reduced. Known as {\it sloppiness}~\cite{Brown_PRE_2003_sloppiness, Waterfall_PRL_2006_sloppiness, Frigerio_arXiv_2024_overcoming_sloppiness}, this phenomenon makes the QFIM anisotropic, i.e., different combinations of encoding operations furnish different SLD-CRBs. Since the displacement and squeezing unitaries do not commute, the SLD-CRB depends on which operation is performed on the probe state first, and thus sloppiness is inevitable in the simultaneous estimation of the respective parameters. 
In terms of the QFIM, the sloppiness is quantified as $s:= \det \mathcal{Q}(\vec{\theta}; \vec{\eta})^{-1}$. For a positive definite QFIM, one can relate it to the SLD-CRB as $s^{1/m} \leq C/m$, where $m$ represents the number of estimable parameters, which, in our case, reads $m = 3$. This means that a lower SLD-CRB corresponds to a lower degree of sloppiness --- an intuitive result since a proper encoding scheme (low sloppiness) would allow for more accurate estimation.

Since the SLD-CRBs for both the ICO scenarios considered in the work never diverge, the QFIM is positive definite, and we can gain some idea about the impact of the ICO process on the sloppiness of the model from $C_{\text{ICO}}^{P(E)_\pm}$. As reported in the previous section, the ICO states with odd parity provide increased precision over the pure encoded Gaussian probe states, implying that ICO helps in mitigating the intrinsic sloppiness arising from the non-commutativity of the encoding operations. More remarkably, although sloppiness arises from the encoding process, ICO during probe preparation can still allow us to overcome it without any change to the encoding process itself. As expected, when we introduce ICO into the encoding process, it has a constructive effect on the sloppiness, leading to lower SLD-CRBs. It will be interesting to study how the ICO scheme benefits the actual measurements used to estimate the parameters, which suffer from incompatibility due to the non-commutativity of the SLDs. An insight into the same may be gained from the trade-off between sloppiness and incompatibility in multiparameter estimation tasks~\cite{He_JPA_2025_scrambling-sloppiness, He_arXiv_2026_fisher-geometry}, but is beyond the scope of this work.

\section{Conclusion}
\label{sec:conclu}

The theory of estimating parameters encoded in quantum systems has garnered unforeseen progress over the past decade. From sensing electric and magnetic fields~\cite{Degen_RMP_2017_sensing_review} to gravimetry~\cite{Armata_PRA_2017_gravimetry_optomechanics, Qvarfort_NC_2018_gravimetry_optomechanics, Szigeti_PRL_2020_gravimetry_BEC}, estimating minuscule leptonic phases~\cite{Ignoti_arXiv_2025_lepton-phase}, and even biological imaging~\cite{Aslam_NRP_2023_bio-imaging}, quantum mechanics can offer substantially higher precision as compared to classical techniques. Beyond single-parameter estimation, simultaneous estimation of multiple parameters, known as multiparameter estimation, has also attracted considerable interest. Moreover, continuous-variable (CV) systems provide promising platforms for implementing quantum information processing tasks, including such metrological tasks, owing to their efficient realisation across a variety of physical platforms.

Focusing on multiparameter estimation in CV systems, we identified the optimal Gaussian probe state for the simultaneous estimation of displacement and squeezing parameters and derived the corresponding SLD-CRB for generic Gaussian probes with real squeezing and displacement along the position and momentum quadratures. At a fixed energy of the probe state, 
we established that the optimal probe is a Gaussian pure state, with both displacement and squeezing along the momentum quadrature. Interestingly, its SLD-CRB decreases with the magnitude of the squeezing amplitude to be estimated, thereby providing higher precision while sensing large squeezing amplitudes. Although the SLD-CRB increases with the probe energy, higher energy permits estimation of larger squeezing strengths and enhances the benefit over generic Gaussian probes. Most importantly, our analysis revealed that simultaneous displacement and squeezing outperforms squeezed vacuum probes having the same energy. Since squeezing is experimentally costly, distributing the available energy between displacement and squeezing offers both lower resource cost and higher estimation precision.

We found that incorporating indefinite causal order (ICO) in probe preparation and encoding, resulting in an odd-superposition probe, can surpass the optimal Gaussian probe, although the advantage is not ubiquitous. Specifically, we showed that ICO in probe preparation provides an advantage over a wider parameter range than ICO in encoding, and the resulting SLD-CRB  could even overcome the lowest error bound achievable by Gaussian states. By considering different configurations of the operations under study, we established that displacement and squeezing along the momentum quadrature were more beneficial than along the position quadrature, akin to the optimal Gaussian probe state. Moreover, the ICO-induced precision enhancement can mitigate the sloppiness inherent in the encoding process arising from the non-commutativity of the encoding operations. Our results further indicate that non-Gaussianity alone does not account for the ICO advantage: the odd-superposition probe can outperform the optimal Gaussian probe even when its relative-entropy-based non-Gaussianity measure vanishes, while the advantage can disappear in parameter regimes where the probe remains non-Gaussian.

Our analysis, therefore, suggests that any advantage from indefinite causal order may arise from the interplay of multiple quantum resources, and hence identifying the features underlying this advantage in quantum multiparameter estimation is a natural direction to explore in the future. It would also be interesting to investigate the Holevo-Cramer-Rao bound under indefinite causal order, as its experimental saturability~\cite{Holevo_2011, Demkowicz-Dobrzaski_JPA_2020_multiparameter_beyond-Fisher, Bressanini_JPA_2024_Gaussian_multiparameter} could provide further practical insights into multiparameter estimation.

\acknowledgements

We acknowledge discussions with Matteo G. A. Paris and Ayan Patra. A.S.D. acknowledges support from the project entitled ``Technology Vertical - Quantum Communication'' under the National Quantum Mission of the Department of Science and Technology (DST)  (Sanction Order No.~DST/QTC/NQM/QComm/$2024/2$ (G)). 
H.S.D. acknowledges financial
support from the Industrial Research \&
Consultancy Centre (IRCC), IIT Bombay via grant (RD$/0521$-IRCCSH$0$-$001$) number $2021289$. 
R.G. acknowledges funding from the HORIZON-EIC-$2022$-PATHFINDERCHALLENGES-$01$ program under Grant Agreement No.~$10111489$ (Veriqub). Views and opinions expressed are those of the authors only and do not necessarily reflect those of the European Union. Neither the European Union nor the granting authority can be held responsible for them.

\appendix

\section{Theory of multiparameter estimation}
\label{app:pre_multiparameter}

In multiparameter quantum estimation, the aim is to estimate a set of parameters, $\vec{\theta} = \{\theta_1, \theta_2, \ldots, \theta_m\}$ as accurately as possible. If the parameters correspond to non-commuting generators, then it has been established that simultaneous estimation can outperform the strategy of estimating each parameter individually~\cite{Humphreys_PRL_2013_multiple-phase_estimation, Baumgratz_PRL_2016_multidimensional-field_estimation}. 
We begin with a probe state $\rho(\vec{\eta})$, characterized by a set of parameters $\vec{\eta}$, which is independent of the parameters to be estimated. The probe state undergoes evolution through a quantum channel, $\Lambda(\vec{\theta})$, which encodes the information about the parameters into the state to yield $\rho(\vec{\theta}; \vec{\eta}) = \Lambda(\vec{\theta})[\rho(\vec{\eta})]$. The encoded state is measured using a POVM, $\hat{\Pi} = \{\hat{\Pi}(x)| \hat{\Pi}(x) \geq 0 ~\forall~ x, \sum_x \hat{\Pi}(x) = \mathbb{I}\}$, whose measurement outcomes are denoted by the set $X=\{x\}$. The probability of obtaining a specific measurement outcome is given by the Born rule as $\mathrm{Pr}(x|\vec{\theta}; \vec{\eta}) = \mathrm{Tr}[\hat{\Pi}(x) \rho(\vec{\theta}; \vec{\eta})]$. If the measurement is performed $N$ times, we obtain a tuple of independent and identically distributed (i.i.d.) outcomes, $\chi = (x_1, x_2, \ldots, x_N)$. 


To accurately determine the encoded parameters, we define an estimator $\check{\theta}_\mathrm{est}(\chi)$, which is essentially maps the set of measurement outcomes, $\chi$, to the set of possible values of the encoded parameters. 
An estimator is considered unbiased if the expectation value of the estimator with respect to the conditional probability distribution approaches the true set of encoded parameters in the limit of a large number of measurements
\begin{eqnarray}
    && \mathbb{E}_{x|\vec{\theta}} [\check{\theta}_\mathrm{est}(\chi)]= \int dx \mathrm{Pr}(x|\vec{\theta}; \vec{\eta}) \check{\theta}_\mathrm{est}(\chi) = \vec{\theta}.
    \label{eq:unbiased estimator}
\end{eqnarray}
It can be demonstrated that for an unbiased estimator, the mean-square error of the estimated parameters is quantified by the elements of the covariance matrix, given by
\begin{eqnarray}
    \nonumber \mathrm{Cov}(\vec{\theta}; \vec{\eta})&=& \mathbb{E}_{x|\vec{\theta}}\left[(\check{\theta}_\mathrm{est}-\mathbb{E}_{x|\vec{\theta}}[\check{\theta}_\mathrm{est}])(\check{\theta}_\mathrm{est}-\mathbb{E}_{x|\vec{\theta}}[\check{\theta}_\mathrm{est}])^T\right],~~~~~
\end{eqnarray}
where $T$ denotes the transposition of the matrix. The covariance matrix contains all the second-order statistical information about the estimation errors. The diagonal elements quantify the statistical error for each parameter, while the off-diagonal elements characterize correlations between estimation errors of different parameters. The covariance matrix is fundamentally bounded by the classical Cramér–Rao inequality~\cite{Rao_1992_CRB, Cramer_PUP_1999_mathematical_statistics}
\begin{equation}
    \mathrm{Cov}(\vec{\theta}; \vec{\eta}) \geq \frac{1}{N} \mathcal{F}\Big(\vec{\theta}, \{\hat{\Pi}_x\}, \vec{\eta} \Big)^{-1}, 
    \label{eq:classical_CRB}
\end{equation}
where $\mathcal{F}(\vec{\theta}, \{\hat{\Pi}_x\}, \vec{\eta})$ is known as the classical Fisher information matrix, which is defined as
\begin{eqnarray}
   \nonumber && \mathcal{F} \Big(\vec{\theta}, \{\hat{\Pi}_x\}, \vec{\eta} \Big)_{ij} = \\
   && \sum_x \mathrm{Pr}(x|\vec{\theta}; \vec{\eta})
   \partial_i [\log \mathrm{Pr}(x|\vec{\theta}; \vec{\eta})]
   \partial_j [\log \mathrm{Pr}(x|\vec{\theta}; \vec{\eta})],~~~~~
\end{eqnarray}
where $\partial_i = \frac{\partial}{\partial \theta_i}$.
While locally unbiased estimators can saturate this bound for any $N$~\cite{Holevo_2011, Suzuki_JPA_2020_nuisance_estimation}, realistic estimators, e.g., the maximum likelihood estimator~\cite{Book_1998_locally_unbiased}, can achieve the same in the asymptotic limit of an infinite number of measurements. 

In the quantum setting, there exist bounds on the covariance matrix which are obtained by maximizing the Fisher information over all possible measurements~\cite{Yuen_IEEE_1973_RLD-CRB, Helstrom_IEEE_1974_non-commuting_detection, Braunstein_PRL_1994_attain-CRB, Braunstein_AP_1996_generalized_uncertainty}. They are thus independent of the measurement POVMs and depend only on the encoded probe state $\rho(\vec{\theta}; \vec{\eta})$. 
The most well-known bound is given by the quantum Fisher information matrix (QFIM)~\cite{Liu_JPA_2020_QFIM_multiparameter}, when defined in terms of the standard logarithmic derivative (SLD) operators, and is known as the SLD-CRB \footnote{The SLD-CRB bounds the trace of the covariance matrix and thus quantifies the joint mean-squared error of all the encoded parameters together.}~\cite{Helstrom_PLA_1967_SLD-CRB, Helstrom_IEEE_1968_SLD-CRB}

\begin{eqnarray}
   && \mathrm{Tr}[\mathrm{Cov}(\vec{\theta}; \vec{\eta})] \geq C(\vec{\theta}; \vec{\eta}) = \mathrm{Tr}[\mathcal{Q}(\vec{\theta}; \vec{\eta})^{-1}], \\
   && \text{with}~ \mathcal{Q}_{ij}(\vec{\theta}; \vec{\eta}) = \frac{1}{2}\,\mathrm{Tr}\!\left[\rho(\vec{\theta}; \vec{\eta}) \left\{ \hat{\mathcal{L}}_i, \hat{\mathcal{L}}_j \right\}_+ \right].
\label{eq_qfim_element}
\end{eqnarray}
Here, $\{A, B\}_+ = AB + BA$ is the anticommutator, $\mathcal{Q}(\vec{\theta}; \vec{\eta})$ is the QFIM, and the SLD operator, $\hat{\mathcal{L}}_i$, corresponding to the parameter $\theta_i$, is implicitly defined as the solution to~\cite{Helstrom_PLA_1967_SLD-CRB, Braunstein_PRL_1994_attain-CRB, Paris_IJQI_2009_metrology-overview}

\begin{eqnarray}
     \frac{d\rho({\vec{\theta}; \vec{\eta}})}{d\theta_i}=\frac{1}{2}\Big( \hat{\mathcal{L}}_i \rho(\vec{\theta}; \vec{\eta}) + \rho(\vec{\theta}; \vec{\eta})\hat{\mathcal{L}}_i \Big).
    \label{eq_rho_sld}
\end{eqnarray}
\noindent For a pure encoded state, the SLD operator can be evaluated directly as
\begin{eqnarray}
   \nonumber \hat{\mathcal{L}}_i=2(|\partial_i\psi(\vec{\theta}; \vec{\eta})\rangle \langle\psi(\vec{\theta}; \vec{\eta})|+|\psi(\vec{\theta}; \vec{\eta})\rangle\langle\partial_i\psi(\vec{\theta}; \vec{\eta})|). \\
    \label{eq_sld_op}
\end{eqnarray}
The SLD-QFIM is real-valued, symmetric, and positive semidefinite, while the SLD operators are Hermitian and traceless. Note that for a single-parameter estimation problem, there exists only one SLD operator. In this case, the QFIM reduces to a scalar quantity $\mathcal{Q}(\theta; \vec{\eta})=\mathrm{Tr}[\rho(\theta; \vec{\eta})\hat{\mathcal{L}}_\theta^2]$. Importantly, in single-parameter quantum estimation, the POVM corresponding to eigenstates of the SLD operator can always be used to physically saturate the SLD-CRB~\cite{Helstrom_JSP_1969_attain-SLD, Braunstein_PRL_1994_attain-CRB, Nagaoka_book_2005_attain-SLD}.

The situation becomes more involved in the multiparameter setting, where more than one SLD operator needs to be considered. In this case, even though the SLD-CRB does not depend on the choice of the POVM, it cannot be saturated unless the SLDs commute with each other, or on average, in which case one can construct a POVM in their common eigenbasis~\cite{Chang_CP_2026_multiparameter_Gaussian}. However, this is not generally the case in realistic physical setups, and due to the noncommutative nature of the SLD operators, it is generally not possible to find a single common eigenbasis in which the POVM elements can be simultaneously defined. Hence, from a measurement perspective, the SLD-CRB is not a tight bound, since it cannot, in general, be saturated, even in the asymptotic limit,
except for pure states where it can always be saturated using collective measurements on an asymptotically large number of copies~\cite{Matsumoto_JPA_2002_matsumoto-bound}. Moreover, a deviation from the SLD-CRB is considered to reveal the quantumness of the quantum statistical model~\cite{Carollo_JSM_2019_multiparameter_quantumness, Razavian_Entropy_2020_multiparameter_quantumness, Candeloro_JPA_2021_asymptotic-incompatibility}. 

The most fundamental scalar bound for multiparameter estimation was proposed by Holevo~\cite{Holevo_JMA_1973_HCRB}. The Holevo-Cramer-Rao bound (HCRB) is tighter than the SLD-CRB by at most a factor of two~\cite{Holevo_2011, Albarelli_PRL_2019, Demkowicz-Dobrzaski_JPA_2020_multiparameter_beyond-Fisher, Zhou_2025} and, for pure states, becomes equal to the NCRB~\cite{Matsumoto_JPA_2002_matsumoto-bound}. While the HCRB requires asymptotic collective measurements to achieve~\cite{Kahn_CMP_2009_HCRB_asymptotic, Yamagata_AS_2013_HRCB_asymptotic, Yang_CMP_2019_HCRB_asymptotic}, it has been demonstrated that using Gaussian probes for displacement sensing~\cite{Holevo_2011} allows for saturating the NCRB even with single-copy measurements.
Calculating the HCRB and the NCRB, however, requires optimization and is generally formulated as a semidefinite programming problem for finite~\cite{Albarelli_PRL_2019, Conlon_NPJQI_2021_NCRB_semidefinite}, as well as infinite-dimensional systems~\cite{Bressanini_JPA_2024_Gaussian_multiparameter, Chang_CP_2026_multiparameter_Gaussian}.

\section{Gaussian states and operations}
\label{app:Gaussian_CV}

Continuous variable systems live in an infinite-dimensional Hilbert space spanned by the eigenstates of the position, $\hat{x}$, and momentum, $\hat{p}$, operators~\cite{Serafini_2017_CV-book}. They can also be equivalently defined by the creation and annihilation operators, $\hat{a}^\dagger = \frac{\hat{x} - \iota \hat{p}}{\sqrt{2}}$ and $\hat{a} = \frac{\hat{x} + \iota \hat{p}}{\sqrt{2}}$ respectively, which satisfy the Bosonic canonical commutation relation (CCR) $[\hat{a}_j, \hat{a}_k^\dagger] = \delta_{jk} ~(\text{or equivalently} [\hat{x}_j, \hat{p}_k] = \iota \delta_{jk})$, where $j, k$ represent independent modes, $\delta_{jk} = 1 ~\text{if}~ j = k ~(0 ~\text{otherwise})$ is the Kronecker delta, and $\iota = \sqrt{-1}$. The creation and annihilation operators, in turn, define the Fock basis, $\{\ket{n}\}$, spanned by eigenstates of the number operator, $\hat{a}^\dagger \hat{a} \ket{n} = n |n\rangle$. Their action on the number eigenstates reads $\hat{a} \ket{n} = \sqrt{n} \ket{n - 1} ( ~\text{with}~ \hat{a} \ket{0} = 0)$ and $\hat{a}^\dagger \ket{n} = \sqrt{n+1} \ket{n+1}$. For an $N$-mode CV state, we can succinctly represent the $2N$ quadrature operators as a vector $\vec{\hat{R}} = (\hat{x}_1, \hat{p}_1, \ldots, \hat{x}_N, \hat{p}_N)^T$, whence the CCR becomes $[\vec{\hat{R}}, \vec{\hat{R}}^T] = \iota \Omega$, where $\Omega$ is the $N$-mode symplectic form defined as $\oplus_{j = 1}^N \omega_j$ with $\omega_j = \begin{pmatrix}
    0 & 1 \\
    -1 & 0
\end{pmatrix}$~\cite{Braunstein_RMP_2005_QI_With_CV}.

One of the principal classes of CV states is the set of Gaussian states~\cite{Ferraro_Bib_2005_Gaussian-CV-review, Adesso_OSID_2014_CV-review}, which can be completely characterized by their first and second moments known as the displacement vector, $\vec{d} = \mathrm{Tr}[\vec{\hat{R}} \rho]$, and the covariance matrix, $\Xi = \mathrm{Tr}[\rho \{(\vec{\hat{R}} - \vec{d}), (\vec{\hat{R}} - \vec{d})^T\}_+]$, respectively. Gaussian operations are those that arise from Hamiltonians at most quadratic in the quadrature operators, and they map the set of Gaussian states to itself. Two paradigmatic Gaussian unitary operations are the displacement operator, $\hat{D}(\alpha) = \exp (\alpha \hat{a}^\dagger - \alpha^* \hat{a})$, and the single-mode squeezing operator, $\hat{S}(z) = \exp \Big( \frac{1}{2}(z \hat{a}^{\dagger 2} - z^* \hat{a}^2) \Big)$. Here, the displacement parameter, $\alpha = \alpha_x + \iota \alpha_p = \mathrm{Tr}[\hat{a} \rho]$, and the squeezing parameter, $z = r e^{\iota \varphi}$, are taken to be complex in general~\cite{Barnett_2002_Optics-book}. The most generic single-mode pure Gaussian state is the displaced squeezed vacuum state, $\ket{\psi (\alpha, z)} = \hat{D}(\alpha) \hat{S} (z) \ket{0}$, which reduces to the coherent state, $\ket{\alpha} = \hat{D}(\alpha) \ket{0}$, and the squeezed state, $\ket{z} = \hat{S}(z) \ket{0}$, when $z = 0$ and $\alpha = 0$ respectively. Note that neither the superposition nor the statistical mixture of multiple Gaussian states results in a Gaussian state, in general. Such states, which have non-vanishing moments beyond the second order, are known as non-Gaussian states~\cite{Walschaers_PRX_2021_non-Gaussian}.

\section{SLD-CRB for generic Gaussian probe states in the multiparameter estimation of displacement and squeezing}
\label{app:SLD_Gaussian}

Recall that we consider two configurations of the probe state, $|{\phi(\vec{\eta})}\rangle_{DS} = \hat{D}(\beta) \hat{S}(\zeta) |0\rangle$ and $|{\phi(\vec{\eta})}\rangle_{SD} = \hat{S}(\zeta) \hat{D}(\beta) |0\rangle$, while the encoding is fixed to $\hat{D}(\alpha) \hat{S}(r)$, with $r, \zeta \in \mathbb{R}^+$, and $\alpha, \beta \in \mathbb{C}$.
In order to calculate the SLD operators as defined in Eq.~\eqref{eq_sld_op}, we first need to perform derivatives with respect to the encoding parameters\footnote{Using $\frac{d}{d \tau} e^{\hat{A}(\tau)} = \hat{A} e^{\hat{A}(\tau)}$ for any operator $\hat{A}(\tau)$ characterized by a parameter $\tau$~\cite{Puri_book_2001_optics}, we obtain the following derivatives for the encoding unitaries
\begin{eqnarray}
   \nonumber \frac{d}{d r} \hat{S}(r) &=& \frac{1}{2} (\hat{a}^{\dagger 2} - \hat{a}^2) \hat{S}(r) \\
   \nonumber \frac{d}{d \alpha_x} \hat{D}(\alpha) &=& (\hat{a}^\dagger - \hat{a} + \iota \alpha_p) \hat{D}(\alpha) \\
   \nonumber \frac{d}{d \alpha_p} \hat{D}(\alpha) &=& \iota (\hat{a}^\dagger + \hat{a} - \alpha_x) \hat{D}(\alpha).
\end{eqnarray}}, 
which, for $|{\phi(\vec{\theta}; \vec{\eta})}\rangle_{DS} = \hat{D}(\alpha) \hat{S}(r) |\phi (\vec{\theta}; \vec{\eta})\rangle_{DS}$,
reads 
\begin{eqnarray}
   \nonumber \partial_{\alpha_x}  
   |{\phi(\vec{\theta}; \vec{\eta})}\rangle_{DS}
   &=& e^{-(\zeta + r)} \ket{e_1} - \iota  (2 e^{-r} \beta_p + \alpha_p) \ket{e_0}, \\
   \label{eq:dphi1ax} \\
  \nonumber \partial_{\alpha_p}  
  |{\phi(\vec{\theta}; \vec{\eta})}\rangle_{DS}
  &=& \iota \Big( e^{\zeta + r} \ket{e_1} + (2 e^r \beta_x + \alpha_x) \ket{e_0} \Big), \\
  \label{eq:dphi1ay} \\
   \nonumber \partial_{r} 
   |{\phi(\vec{\theta}; \vec{\eta})}\rangle_{DS}
   &=& \frac{1}{2} \Big( (\beta^{*2} - \beta^2) \ket{e_0} + \sqrt{2} \ket{e_2} \\
   \nonumber &+& (2 \beta^* \cosh \zeta - 2 \beta \sinh \zeta) \ket{e_1} \Big). \\
   \label{eq:dphi1r}
\end{eqnarray}
Here, for brevity, we have defined $\ket{e_n} = \hat{D}(\alpha) \hat{S}(r) \hat{D}(\beta) \hat{S}(\xi) \ket{n}$. A similar analysis yields the derivatives for 
$|{\phi(\vec{\theta}; \vec{\eta})}\rangle_{SD} = \hat{D}(\alpha) \hat{S}(r) |\phi (\vec{\theta}; \vec{\eta})\rangle_{SD}$ as

\begin{eqnarray}
  \nonumber \partial_{\alpha_x}  
  |{\phi_2(\vec{\theta}; \vec{\eta})}\rangle_{SD}
  &=& e^{-(\zeta + r)} \ket{f_1} \\
  &-& \iota (2 e^{-r - \zeta} \beta_p + \alpha_p) \ket{f_0}, 
  \label{eq:dphi2ax}\\
 \nonumber  \partial_{\alpha_p}  
 |{\phi_2(\vec{\theta}; \vec{\eta})}\rangle_{SD}
 &=& \iota \Big( e^{\zeta + r} \ket{f_1} + (2 e^{r + \zeta} \beta_x + \alpha_x) \ket{f_0} \Big), \\
  \label{eq:dphi2ay}\\
   \nonumber \partial_{r} 
   |{\phi_2(\vec{\theta}; \vec{\eta})}\rangle_{SD}
   &=& \frac{1}{2} \Big( (\beta^{*2} - \beta^2) \ket{f_0} + 2 \beta^* \ket{f_1} \\
   &+& \sqrt{2} \ket{f_2} \Big),
   \label{eq:dphi2r}
\end{eqnarray}
where $\ket{f_n} = \hat{D}(\alpha) \hat{S}(r) \hat{S}(\xi)  \hat{D}(\beta) \ket{n}$. Note that both $\{\ket{e_n}\}_{n = 0}^{\infty}$ and $\{\ket{f_n}\}_{n = 0}^{\infty}$ form an orthonormal basis satisfying the completeness relation $\sum_{n = 0}^{\infty} \ketbra{e_n}{e_n} = \sum_{n = 0}^{\infty} \ketbra{f_n}{f_n} = \mathbb{I}$.
Using Eqs.~\eqref{eq:dphi1ax} - ~\eqref{eq:dphi2r} in Eq.~\eqref{eq_sld_op}, we can construct the QFIM for both the probe states, from which follow their respective SLD-CRBs

\begin{eqnarray}
    \nonumber C^{P}_{DS}(\vec{\theta}; \vec{\eta}) &=& \cosh^2 (r \pm \zeta) + \frac{e^{2r}\beta_x^2 + e^{-2r}\beta_p^2}{2} \\
    && \mathrm{for} |\phi(\vec{\theta}; \vec{\eta})\rangle_{DS} ~\mathrm{and}~ \label{eq:app_Gaussian_sld} \\
    \nonumber C^{P}_{SD}(\vec{\theta};\vec{\eta})&=& \cosh^2 (r \pm \zeta) + \frac{e^{2(r \pm \zeta)}\beta_x^2 + e^{-2(r \pm \zeta)}\beta_p^2}{2} \\
    && \mathrm{for} |\phi(\vec{\theta}; \vec{\eta})\rangle_{SD} \label{eq:app_Gaussian_sld_reverse_order},
\end{eqnarray}
where $\pm \zeta$ corresponds to position and momentum squeezing in the probe preparation, respectively.

\section{Optimal pure single-mode Gaussian probe state for multiparameter sensing of displacement and squeezing parameters}
\label{app:optimal}

In this appendix, we investigate the optimal Gaussian probe state that minimizes both the SLD-CRBs in Eqs.~\eqref{eq:app_Gaussian_sld} and~\eqref{eq:app_Gaussian_sld_reverse_order}, when the energy of the probe state is fixed.
Recall that the energy of the considered Gaussian probe states are $\mathcal{E}^{P}_{DS} = \sinh^2{\zeta} + \beta_x^2 + \beta_p^2$ and $\mathcal{E}^{P}_{SD} = \sinh^2{\zeta} + e^{\pm \zeta} \beta_x^2 + e^{\mp \zeta} \beta_p^2$ (since the encoding parameters are unknown, we do not consider the energy of the encoding step). 
To tackle the minimization of $C^{P}_{DS}(C^{P}_{SD})$, with respect to the probe parameters, $\vec{\eta} = \{\beta_x, \beta_p, \pm \zeta \}$, at a fixed $\mathcal{E}^{P}_{DS} (\mathcal{E}^{P}_{SD}) = \mathcal{E}^{P}$, we resort to the theory of Lagrange undetermined multipliers. Defining $L_{DS} = C^{P}_{DS} - \lambda \mathcal{E}^{P}_{DS}$ and $L_{SD} = C^{P}_{SD} - \lambda \mathcal{E}^{P}_{SD}$, with $\lambda$ being the undetermined multiplier, we determine the following derivatives with respect to the probe parameters

\begin{eqnarray}
\frac{\partial L_{DS}}{\partial \beta_x} &=& \beta_x(e^{2r} - 2\lambda); ~ \frac{\partial L_{SD}}{\partial \beta_x} = e^{\pm \zeta}\beta_x(e^{2r} - 2\lambda),
\label{eq:constraints1} \\
\frac{\partial L_{DS}}{\partial \beta_p} &=& \beta_p(e^{-2r} - 2\lambda); ~ \frac{\partial L_{SD}}{\partial \beta_p} = e^{\mp \zeta} \beta_p(e^{-2r} - 2\lambda),~~~~~~
\label{eq:constraints2} \\
\frac{\partial L_{DS}}{\partial \zeta}
&=& \pm \sinh 2 (r \pm \zeta) - \lambda \sinh 2 \zeta,
\label{eq:constraints3} \\
\nonumber \frac{\partial L_{SD}}{\partial \zeta}
&=& \pm \sinh 2 (r \pm \zeta) - \lambda \sinh 2 \zeta \\
\nonumber && \mp \beta_x^2 e^{\mp 2 \zeta}(e^{2r} - 2 \lambda) \mp \beta_p^2 e^{-2(r \mp \zeta)}(2 \lambda e^{2r} - 1). \\
\label{eq:constraints3a} 
\end{eqnarray}
Note that the derivatives with respect to $\lambda$ recover the energy constraint and ensure that we consider probes of fixed energy, $\mathcal{E}^P$.
The Eqs. ~\eqref{eq:constraints1} and~\eqref{eq:constraints2} respectively vanish when either $(1).~ \beta_x = 0 ~\text{or}~ (e^{2r} - 2 \lambda) = 0$ and $(2).~ \beta_p = 0 ~\text{or}~ (e^{-2r} - 2 \lambda) = 0$. Let us consider the two cases one by one. 

$\mathbf{1.}$ Setting $\beta_x \neq 0$ yields $\lambda = e^{2r}/2$, and substituting in Eq.~\eqref{eq:constraints3} we obtain $\zeta^{\pm}_\mathrm{opt} = \pm \log (2 e^{-4r} - 1)/4$, where $+(-)$ again indicates position- (momentum)-squeezed probes. It is immediately apparent that, for position-squeezed probes, the optimal squeezing strength is negative (except in the trivial case of $r = 0$), which is unphysical. For momentum squeezing, $\zeta^{-}_\mathrm{opt} \in \mathbb{R}^+$ only when $r < \frac{1}{4} \log2$ such that $(2e^{-4r} - 1) > 1$. Since the value of the optimal probe squeezing strength is not defined for all possible values of the encoding squeezing strength, we will not consider it as an optimal solution. Therefore, we must have $\beta_x = 0$ for the optimal probe.

$\mathbf{2.}$ If we consider $\beta_p \neq 0$, the undetermined multiplier assumes the form $\lambda = e^{-2r}/2$. Using Eq.~\eqref{eq:constraints3}, the optimal probe-squeezing strength becomes $\zeta^{\pm}_\mathrm{opt} =  \mp \log (2 e^{4r} - 1)/4$, 
which is again non-positive for the case of position-squeezing, and hence, unphysical. On the other hand, for momentum-squeezed probes, $\zeta^{-}_\mathrm{opt} = \log (2 e^{4r} - 1)/4 \geq 0 ~\forall~ r$. 

We thus arrive at the optimal single-mode Gaussian probe state for our protocol, which has been analyzed in detail in the main text.\\

\section{SLDs for multiparameter estimation using ICO}
\label{app:SLD_ICO}

In this appendix, we derive the SLDs required for computing the SLD-CRB for multiparameter estimation using the ICO protocol.

\subsection*{ICO during probe preparation}

We first consider keeping the encoding order of operations fixed to $\Lambda(\vec{\theta}) = \mathcal{D} \circ \mathcal{S}$, which, however, is applied on $\ket{\phi(\vec{\eta})}_{\pm}$, the probe state prepared using the quantum switch

\begin{eqnarray}
   \nonumber \ket{\phi(\vec{\eta})}_{\pm} \sim \Big( \sqrt{\frac{p}{2}} \hat{D}(\beta) \hat{S}(\xi) \pm \sqrt{\frac{1 - p}{2}} \hat{S}(\xi) \hat{D}(\beta) \Big) \ket{0}, \\
\end{eqnarray}
where $\xi \in \{ \pm \zeta \}$ as before, and its positive and negative values indicate position and momentum squeezing, respectively. Note that we have not specified the normalization constants, $N_\pm$, for the two states in Eq.~\eqref{eq:ICO_probe}. We shall refer to the two configurations, $\ket{\phi(\vec{\eta})}_{+}$ and $\ket{\phi(\vec{\eta})}_{-}$, as even and odd superpositions, respectively. The normalisation factors are given as

\begin{eqnarray}
   \nonumber N_\pm &=& \frac{1}{2} \pm \sqrt{p(1 - p)} \cos \Big( 2 \beta_x  \beta_p \sinh \xi \Big) \\
   &+& \nonumber \exp \Big( - \frac{1}{2} e^{2\xi} (e^{\xi} - 1)^2 \times \\
   && ~~~~~~~(\beta_x^2 + e^{2 \xi} \beta_p^2) \Big). 
\end{eqnarray}

Some cumbersome algebra allows us to derive the SLDs from Eq.~\eqref{eq_sld_op} using

\begin{eqnarray}
  && \nonumber \partial_{\alpha_x} |\phi(\vec{\theta}; \vec{\eta})\rangle_{\pm} = \\
  && \nonumber \frac{1}{\sqrt{N_\pm}} \Bigg[ \sqrt{\frac{p}{2}} \Big( e^{-(r + \xi)} \ket{e_1} - \iota (2 e^{-r} \beta_p + \alpha_p) \ket{e_0} \Big) \\
    && \nonumber ~~~~~~~ \pm \sqrt{\frac{1 - p}{2}} \Big( e^{-(r + \xi)} \ket{f_1} - \\
    && \nonumber ~~~~~~~~~~~~~~~~~~~~~~~\iota (2 e^{-(r + \xi )} \beta_p + \alpha_p) \ket{f_0} \Big) \Bigg], \\ 
    \label{eq:ICO_probe_ax}
\end{eqnarray}
\begin{eqnarray}
    && \nonumber \partial_{\alpha_p} |\phi(\vec{\theta}; \vec{\eta})\rangle_{\pm} = \\
    && \nonumber \frac{\iota}{\sqrt{N_\pm}} \Bigg[ \sqrt{\frac{p}{2}} \Big( e^{(r + \xi)} \ket{e_1} + (2 e^{r} \beta_x + \alpha_x) \ket{e_0} \Big)  \\
    && \nonumber ~~~~~~~\pm  \sqrt{\frac{1 - p}{2}} \Big( e^{(r + \xi )} \ket{f_1} + \\
    && \nonumber ~~~~~~~~~~~~~~~~~~~~~~~(2 e^{(r + \xi)} \beta_x + \alpha_x) \ket{f_0} \Big) \Bigg], \\ 
    \label{eq:ICO_probe_ay} 
    \end{eqnarray}
\begin{eqnarray}
        && \nonumber \partial_{r} |\phi(\vec{\theta}; \vec{\eta})\rangle_{\pm} = \\
     && \nonumber \frac{1}{2 \sqrt{N_\pm}} \Bigg[ \sqrt{\frac{p}{2}} \Big(2( \beta^* \cosh \xi + \beta \sinh \xi) \ket{e_1} \\
     && \nonumber  ~~~~~~~~~~~~~~~~~~~-4 \iota \beta_x \beta_p \ket{e_0} + \sqrt{2} \ket{e_2} \Big) \\
     && \nonumber ~~~~~~~~\pm \sqrt{\frac{1 - p}{2}} \Big( 2 \beta^* \ket{f_1} -4 \iota \beta_x \beta_p \ket{f_0} + \sqrt{2} \ket{f_2} \Big) \Bigg]. \\
     \label{eq:ICO_probe_r}
     \end{eqnarray}
As before, we define two orthonormal sets of states, $\{\ket{e_n} = \hat{D}(\alpha) \hat{S}(r) \hat{D}(\beta) \hat{S}(\xi) \ket{n} \}$ and $\{\ket{f_n} = \hat{D}(\alpha) \hat{S}(r) \hat{S}(\xi) \hat{D}(\beta) \ket{n} \}$. 

\subsection*{Encoding through ICO}

Next, we consider ICO in the encoding process. The unnormalized encoded states corresponding to the two measurement outcomes are 

\begin{eqnarray}
   \nonumber |\tilde{\phi}(\vec{\theta}; \vec{\eta})\rangle_\pm &\sim& \sqrt{\frac{p}{2}} \hat{D}(\alpha) \hat{S}(r) \ket{\phi(\vec{\eta})}_{DS} \\
   &\pm& \sqrt{\frac{1 - p}{2}} \hat{S}(r) \hat{D}(\alpha) \ket{\phi(\vec{\eta})}_{DS}.
   \label{eq:ICO_encoding_state}
\end{eqnarray}
Note that since the encoding itself is done through ICO, the normalisation term of the encoded state is also a function of the parameters to be estimated, i.e., $N_{\pm}(\vec{\theta} = \{ \alpha, r \} )$, and thus must also be taken into account while computing the SLDs. Specifically, the derivative of the encoded state with respect to a given parameter, say $\theta_i$, reads

\begin{eqnarray}
   \nonumber && \partial_{\theta_i} \Bigg( \frac{1}{\sqrt{N_\pm(\vec{\theta})}} \ket{\tilde{\phi}(\vec{\theta}; \vec{\eta})}_\pm \Bigg) = \partial_{\theta_i} \Bigg( \frac{1}{\sqrt{N_\pm(\vec{\theta})}} \Bigg) \ket{\tilde{\phi}(\vec{\theta}; \vec{\eta})}_\pm \\
  \nonumber && ~~~~~~~~~~~~~~~~~~~~~~~~~~~~~~~~~ +  \frac{1}{\sqrt{N_\pm(\vec{\theta})}} \partial_{\theta_i} \Bigg( \ket{\tilde{\phi}(\vec{\theta}; \vec{\eta})}_\pm \Bigg), \\
  \label{eq:ICO_encoding_derivative1}
\end{eqnarray}
where, using $N_\pm(\vec{\theta}) = \langle \tilde{\phi}(\vec{\theta}; \vec{\eta}) | \tilde{\phi}(\vec{\theta}; \vec{\eta}) \rangle_\pm$, we can evaluate its derivative in terms of those of the unnormalized state as

\begin{eqnarray}
   \nonumber &&  \partial_{\theta_i} \Bigg( \frac{1}{\sqrt{N_\pm(\vec{\theta})}} \Bigg) = - \frac{1}{2} N_{\pm}^{-3/2}(\vec{\theta}) \Big( \langle \tilde{\phi}(\vec{\theta}; \vec{\eta}) | \partial_{\theta_i} \tilde{\phi}(\vec{\theta}; \vec{\eta}) \rangle_\pm \\
   \nonumber && ~~~~~~~~~~~~~~~~~~~~~~~~~~~~~~~~~~~~~~~~+  \langle \partial_{\theta_i} \tilde{\phi}(\vec{\theta}; \vec{\eta}) | \tilde{\phi}(\vec{\theta}; \vec{\eta}) \rangle_\pm  \Big). \\
   \label{eq:ICO_encoding_derivative2}
\end{eqnarray}
Finally, defining $\ket{\tilde{e}_n} = \hat{D}(\alpha) \hat{S}(r) \hat{D}(\beta) \hat{S}(\xi) \ket{n}$ and $\ket{\tilde{f}_n} = \hat{S}(r) \hat{D}(\alpha) \hat{D}(\beta) \hat{S}(\xi) \ket{n}$, the derivatives of the unnormalized encoded probe with respect to the estimable parameters read

\begin{eqnarray}
    && \nonumber \partial_{\alpha_x} |\tilde{\phi}(\vec{\theta}; \vec{\eta})\rangle_{\pm} = \\
  && \nonumber  \sqrt{\frac{p}{2}} \Big( e^{-(r \pm \zeta)} \ket{e_1} - \iota (2 e^{-r} \beta_p + \alpha_p) \ket{e_0} \Big) \\
    && \nonumber ~~~~~~~ \pm \sqrt{\frac{1 - p}{2}} \Big( e^{\mp \zeta} \ket{f_1} - \\
    && \nonumber ~~~~~~~~~~~~~~~~~~~~~~~ \iota (2  \beta_p + \alpha_p) \ket{f_0} \Big) , \\ 
   \label{eq:ICO_encoding_ax}
   \end{eqnarray}
   \begin{eqnarray}
    && \nonumber \partial_{\alpha_p} |\tilde{\phi}(\vec{\theta}; \vec{\eta})\rangle_{\pm} = \\
    && \nonumber  \iota \sqrt{\frac{p}{2}} \Big( e^{(r \pm \zeta)} \ket{e_1} + (2 e^{r} \beta_x + \alpha_x) \ket{e_0} \Big)  \\
    && \nonumber ~~~~~~~\pm \iota \sqrt{\frac{1 - p}{2}} \Big( e^{\pm \zeta} \ket{f_1} + \\
    && \nonumber ~~~~~~~~~~~~~~~~~~~~~~~(2 \beta_x + \alpha_x) \ket{f_0} \Big) , \\ 
   \label{eq:ICO_encoding_ay}
   \end{eqnarray}
   \begin{eqnarray}
        && \nonumber \partial_{r} |\tilde{\phi}(\vec{\theta}; \vec{\eta})\rangle_{\pm} = \\
     && \nonumber \frac{1}{2} \sqrt{\frac{p}{2}} \Big( 2 ( \beta^* \cosh \zeta \mp \beta \sinh \zeta) \ket{e_1} \\
     && \nonumber  ~~~~~~~~~~~~~~~~~~~-4 \iota \beta_x \beta_p \ket{e_0} + \sqrt{2} \ket{e_2} \Big) \\
     && \nonumber ~~~~~~~~\pm \frac{1}{2} \sqrt{\frac{1 - p}{2}} \Bigg( \sqrt{2} \ket{f_2} - 4 \iota (\alpha_x + \beta_x) (\alpha_p + \beta_p) \ket{f_0} \\
    \nonumber && ~~~~~~~~~~~~~~~~~~~~~~~~~~~~ + 2 \Big( e^{\mp \zeta} (\alpha_x + \beta_x) \\
    && ~~~~~~~~~~~~~~~~~~~~~~~~~~~~ - \iota e^{\pm \zeta} (\alpha_p + \beta_p) \Big) \ket{f_1}  \Bigg).
    \label{eq:ICO_encoding_r}
\end{eqnarray}

\bibliography{ref}

\end{document}